\documentclass{aa}  

\usepackage{graphicx}
\usepackage{txfonts}
\usepackage{hyperref}
\usepackage{color}

\begin{document}

   \title{Linear acceleration emission of pulsar relativistic streaming instability and interacting plasma bunches}

   \titlerunning{Pulsar linear acceleration emission}

   \author{Jan Ben\'{a}\v{c}ek
       \inst{1,2} \and
          Patricio~A.~Mu\~noz
          \inst{3,2} \and
          J\"org~B\"uchner
          \inst{2,3} \and
          Axel~Jessner\inst{4}
          }
          
    \authorrunning{Benáček et al.}

   \institute{
        Now at: Institute for Physics and Astronomy, University of Potsdam, D-14476 Potsdam, Germany \\ \email{jan.benacek@uni-potsdam.de}
        \and
                Center for Astronomy and Astrophysics, Technical University of Berlin, D-10623 Berlin,
                Germany 
                \and
                Max Planck Institute for Solar System Research, D-37077 G\"ottingen, Germany
         \and
         Max-Planck Institute for Radio Astronomy, D-53121 Bonn, Germany \\
             }

   \date{Received: ; accepted:}

 
  \abstract
   {Linear acceleration emission is one of the mechanisms that might explain intense coherent emissions of radio pulsars.
        This mechanism is not well understood, however, because the effects of collective plasma response and nonlinear plasma evolution on the resulting emission power must be taken into account. 
    In addition, details of the radio emission properties of this mechanism are unknown, which limits the observational verification of the emission model.
   }
   {By including collective and nonlinear plasma effects, we calculate radio emission power properties by the linear acceleration emission mechanism that occurs via the antenna principle for two instabilities in neutron star magnetospheres: 1) the relativistic streaming instability, and 2) interactions of plasma bunches.
   }
   {We used 1D electrostatic relativistic particle-in-cell simulations to evolve the instabilities self-consistently.
   From the simulations, the power properties of coherent emission were obtained by novel postprocessing of electric currents. 
   }
   {We found that the total radio power by plasma bunch interactions exceeds the power of the streaming instability by eight orders of magnitude.
    The wave power generated by a plasma bunch interaction can be as large as $2.6\times10^{16}$~W.
    The number of bunch interactions that are required to explain the typical pulsar power, $10^{18}$--$10^{22}$~W, depends on how the coherent emissions of bunches are added up together.
    Although $\sim$$4\times (10^1-10^5)$ simultaneously emitting bunches are necessary for an incoherent addition of their radiation power, $\gtrsim 6-600$ bunches can explain the total pulsar power if they add up coherently.
        The radio spectrum of the plasma bunch is characterized by a flatter profile for low frequencies and by a power-law index up to $\approx-1.6 \pm 0.2$ for high frequencies.
    The plasma bunches simultaneously radiate in a wide range of frequencies, fulfilling no specific relation between emission frequency and height in the magnetosphere.
The power of the streaming instability is more narrowband than that of the interacting bunches, with a high-frequency cutoff.
In both instabilities, the angular width of the radiation decreases with increasing frequency.  
In addition, the wave power evolution depends on the pulsar rotation angle, causing microsecond fluctuations in the intensity because it oscillates between positive and negative wave interference as a function of the emission angle.
   }
   {
   }

   \keywords{ pulsars: general -- Stars: neutron -- Plasmas -- Instabilities
               }

   \maketitle
%

\section{Introduction} \label{sec:intro}
Pulsars are strongly magnetized neutron stars that emit coherent radio waves \citep{Sturrock1971,Ruderman1975,Beskin1993}.
For more than 50 years, the nature of their coherent radio emission from relativistic plasma in their magnetospheres as well as the exact emission mechanism have been discussed \citep{Weatherall1997,Melrose1999,Michel2004,Eilek2016,Beskin2018,Melrose2020a}.


Most of the current pulsar magnetospheric models rely on the concept of Goldreich--Julian currents \citep{Goldreich1969} and sparking events that form electron--positron pairs in polar cap regions or magnetospheric current sheets \citep{Ruderman1975,Cheng1977b,Cheng1977a,Buschauer1977,Chen2014,Cerutti2015,Philippov2022}.
They assume a strong electric field component ($E \approx 10^{12}\,\mathrm{V}\,\mathrm{m}^{-1}$) directed parallel to the local magnetic field ($\boldsymbol{E}\cdot\boldsymbol{B}\neq0$), formed in gap regions in open magnetic field lines along which particles can escape the magnetosphere.
If the current densities are low in the gaps, the currents do not fully screen the convective electric fields so that particles can be accelerated to ultrarelativistic velocities.
Particles with typical Lorentz factors $\gamma = 10^{6}-10^{8}$ form a primary beam.
During the acceleration, the particles emit $\gamma$-ray photons that can propagate in the magnetosphere.
In strong pulsar magnetic fields, the photons decay into electron--positron pairs and form a secondary beam. The secondary particles typically outnumber primary particles by $10^{3}-10^{5}$ times \citep{Timokhin2019}.
The Lorentz factors of the produced secondary particles are in the range $10^2$--$10^4$ \citep{Arendt2002}.

Linear acceleration emission (LAE) is one of the mechanisms that has been proposed to explain various types of pulsar radiation \citep{Cocke1973,Ginzburg1975,Melrose1978,Kroll1979,Rowe1992a,Rowe1992b}.
The emission mechanism assumes particles (e.g., the primary or secondary particles)
that undergo acceleration parallel to the magnetic field.
The most simple model assumes that the particles are accelerated in electrostatic waves in gap regions and emit electromagnetic waves.

The LAE mechanism is thought apply in pulsars according to four main cases \citep{Melrose2009b} outlining four principal scenarios in which the LAE mechanism may become important for pulsars.

In the first scenario, a coherent LAE mechanism requires electrostatic waves with smaller amplitude (compared to the cases below). 
In these waves, particles oscillate with a velocity amplitude $v_\mathrm{max}$. 
The maximum Lorentz factor of the oscillations, $\gamma_\mathrm{max} = (1 - v_\mathrm{max}^2/c^2)^{-\frac{1}{2}}$, should be $\gamma_\mathrm{max} \lesssim 10$ to maintain the emitted waves coherent \citep{Melrose2009a}.
Because the amplitude of the particle velocity is relativistic, the typical emission frequency therefore is higher than the oscillation frequency.
                
Two versions of the coherent mechanism exist:
In the antenna mechanism, the coherence is provided by particle grouping and particles that emit in phase \citep{Ginzburg1975,Benford1977}.
As the phases of individual particles are not random, their individual contributions are summed up. In this way, their radiation intensity exceeds the sum of incoherent intensities that occurs when particles emit randomly.
The emitted waves propagate without additional amplification.
Because the emission frequency of the relativistic particles is higher than the oscillation frequency, which may be the local relativistic plasma frequency, for instance, the produced electromagnetic waves can propagate through the plasma.
                
In the maser mechanism of the coherent LAE the emitted waves as they propagate through the plasma are amplified \citep{Melrose1978}.
The amplification can occur, for example, by an inverse population of energy states of particles that has to be created and maintained, implying a negative absorption coefficient.
In comparison with the antenna mechanism, no preliminary phasing or grouping of particles is necessary.

The maser mechanism can be considered similar to the klystron radiation mechanism \citep{Rylov1978}.
In a klystron, a beam of charged particles with a given initial velocity distribution is injected into an extended longitudinal region.
The velocity of the beam particles is modulated by a varying electric field.
The particles continue to drift with their imprinted velocities, which converts the velocity modulation into a charge density modulation.
The particles may then amplify the initial electromagnetic fields. 
        
Second, LAE can be a high-energy emission process in a strong electric field, for instance, by primary particles in the gap region. The released photons have an energy of several tens of keV \citep{Akhiezer1975,Levinson2005}. 

Third, LAE can provide a source of secondary pairs if the emission mechanism produces high-energy $\gamma$-ray photons that can decay into electron--positron pairs. This requires a minimum energy of the $\gamma$-ray photons of the order of MeV \citep{Luo2008,Philippov2020,Cruz2020}.

Fourth, LAE can contribute to a damping of large-amplitude electrostatic waves.
The waves appear during sparks in the polar cap region \citep{Melrose2009b}.

In recent years, LAE power rates have been calculated in various test-particle approaches that were then extrapolated to represent the emitting particles in the plasma \citep{Luo2008,Melrose2009a,Melrose2009b,Reville2010}.
The question of how the LAE works in a plasma in which collective particle effects are intrinsic and feedback effects on the electromagnetic field are essential has not yet been treated for pulsars.
In addition to pulsars, the LAE mechanism can also apply to black hole magnetospheres \citep{Levinson2005,Levinson2018} or to the interaction of plasma bunches in models of fast radio bursts \citep{Lu2018,Yang2020,Zhang2020}.

This paper focuses on the coherent version of the LAE mechanism that occurs based on the antenna principle, that is, case~1(a) above.
To the best of our knowledge, the emission properties of the LAE in terms of the angular profile, radio spectra, temporal evolution, or wave interference have not been analyzed in detail.
In particular, the radio emission of the LAE mechanism has not been investigated in plasma kinetic simulations in which particles and waves evolve self-consistently.
The advantage of the kinetic simulations is that they provide the necessary information about nonlinear and collective plasma evolution so that the properties of the coherent LAE mechanism can be estimated.

We computed the LAE radio power of two plasma instabilities produced by pulsar gap regions.
(1) The streaming instability can be produced by overlapping plasma bunches \citep{Buschauer1977,Usov1987,Weatherall1994,Rahaman2020}.
We found that the bunches can overlap and form a streaming instability if there is no initial drift velocity between electrons and positrons \citep{Manthei2021,Benacek2021a}.
(2) Interactions of plasma bunches can have a nonzero drift velocity between electrons and positrons.
The drift velocity can be obtained when the electron--positron pairs are created in the electric field of a gap region and are accelerated in opposite directions in these electric fields  \citep{Levinson2005,Timokhin2010,Timokhin2013,Benacek2021b}.

This paper is structured as follows.
We discuss our approach for calculating LAE, describe how we considered the wave coherence, and present our calculation steps in Sect.~\ref{methods}.
In Sect.~\ref{sec:results} we present the results of LAE simulations of the streaming instability and the bunch interaction model.
We discuss the relevance of the LAE mechanism for pulsar radio emissions \ref{sec:discussion}.
Section~\ref{sec:conclusion} includes our conclusions.
Details and the numerical implementation of the LAE calculation steps and tests are summarized in Appendices~\ref{methods1}, \ref{methods2}, and \ref{methods3}.

\section{Methods} \label{methods}
\subsection{Geometry of the emission}

\begin{figure*}
    \centering
        \includegraphics[width=0.7\textwidth]{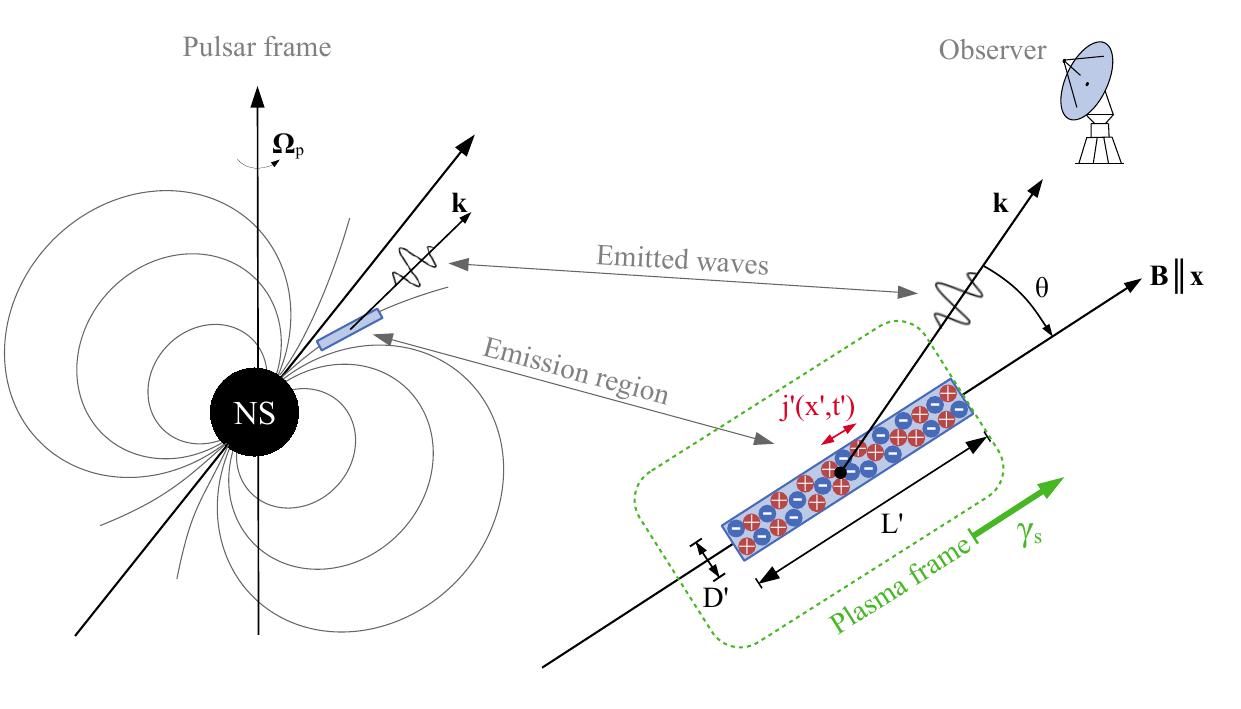}
        \caption{
            Scheme of the considered emission region in the open magnetic field lines of a neutron star magnetosphere.
            The emission region (plasma frame) moves in the pulsar frame with a Lorentz factor $\gamma_\mathrm{s}$ along the magnetic field line.
            In our approach, we considered the electric current density $j'(\vec{x'},t')$ as a function of space and time.
            The plasma frame is denoted by primes everywhere except 
in Appendices~\ref{methods1} and \ref{methods2}.
            In the plasma frame, the emission region has a cylindrical shape with a length $L'$ along the axis $\vec{x}' \parallel \vec{B}$ and a diameter $D' \ll L'$. 
        The length $L'$ depends on simulation parameters (see text), but we have fixed the diameter to notional $D' = 1.13$\,m to calculate the emission power.
            In the pulsar reference frame, the radio emission occurs at a wave vector $\vec{k} = k \cdot \vec{n} = \omega / c \cdot \vec{n}$ at an angle $\theta$ to the magnetic field. 
        }
        \label{fig:emission-scheme}
\end{figure*}

The scheme of the considered radio emission region is presented in Fig.~\ref{fig:emission-scheme}.
We assumed that plasma bunches move along open magnetic field lines in the magnetosphere of a neutron star.
In this paper, we focus on a model for a localized fundamental emission process that can be the source of the pulsar radio emission.
The emission is likely to be affected by propagation through an inhomogeneous pulsar magnetosphere in a time- and frequency-dependent manner. 
However, we did not attempt to model a global complex and time-variable magnetosphere and how it affects the propagation of broadband LAE radio waves generated inside it. 
Instead, we provide estimates of spectra and pulse shapes as if they had escaped unchanged in the direction of the observer.

The emission region of the bunches produces radio waves by a coherent LAE mechanism.
In the comoving plasma reference frame (denoted with primes), the emission region is assumed to have a cylindrical shape characterized by its length $L'$ along the magnetic field line and its diameter $D'$.
 The diameter $D'$ of the cylinder is considered smaller than the wavelength of the emitted waves $\lambda$.
Because the emission region is considered much longer $L'$ than its diameter $D'$, the emission region is similar to an emitting antenna.

In the emission region, a nonzero oscillating electric current $j'(\vec{x}',t')$ varies only along the cylinder, that is, along the magnetic field line and $\vec{x}$-axis.
The perpendicular profile of the current was assumed to be uniform inside $D'$ because of its small size and zero outside.
Moreover, the perpendicular component of the electric current was neglected, $j_\perp'(\vec{x}) = 0$, because the plasma particles were confined to only move along the magnetic field lines.

If a spatial element of the current oscillates, plasma particles associated with the oscillations can emit electromagnetic radio waves.
Emitted electromagnetic waves by individual current elements were added up along $\vec{x}$ according to their wave phases and emission angle.
Each emitted wave propagated into a direction given by a wave vector $\vec{k}'$ in the plasma frame.
At large distances from the emission region, the size of the emission region can be neglected, and the wave vector of superposed electromagnetic waves can be described in spherical coordinates $(r, \theta, \varphi)$.
In these coordinates, the radiation pattern is symmetric in azimuthal angle $\varphi$, and the wave vector has a polar angle $\theta$ to the magnetic field.

The emitting plasma region moves in the pulsar reference frame with a Lorentz factor $\gamma_\mathrm{s} = (1 - \beta_\mathrm{s}^2)^{-\frac{1}{2}}$, assuming $\beta_\mathrm{s} > 0$, on a trajectory along the field lines in the radiation direction.
The emission region rotates with an angular frequency $\Omega_\mathrm{p} = 2\pi / T_\mathrm{pulsar}$, where $T_\mathrm{pulsar}$ is the pulsar period.
The velocity with respect to the observer is then the relativistic addition of the angular velocity $v_\mathrm{rot} = R \Omega_\mathrm{p} \cos(\theta)$, where $R$ is the distance from the pulsar rotational axis, with the longitudinal velocity $\beta_\mathrm{s}c$, and we assumed that the resulting Lorentz factor was $\gamma_s=100$. 
Hence, for each emission angle $\theta'$ in the plasma frame, the Lorentz transformation results in an emission angle $\theta$ in the pulsar (observer) frame.
Furthermore, the size of the emission region as seen by an observer in the pulsar frame varies because the relativistic Doppler effects depend on $\theta$.
For example, if the emission region emits radiation at an arbitrary frequency in the plasma reference frame, the observed frequency in the pulsar frame is highest for $\theta = 0$, decreases with increasing $\theta$, and is lowest for $\theta = \pi$.

As the neutron star rotates with an angular frequency $\Omega_\mathrm{p}$, the observer detects the radiation with changing angle $\theta$.
Therefore, the detected emission power as a function of $\theta$ can be converted into changes in time $t$.

\subsection{Considering the wave coherence} 
The standard approach to obtain LAE is tracking individual plasma particles to obtain the emission power incoherently  (e.g., \citet{Nishikawa2021}).
However, the coherent approach of the mechanism requires taking collective particle motion into account and adding emitted waves according to their phases.

We calculated the wave emission properties directly from the aggregated electric currents in 1D kinetic particle-in-cell (PIC) simulations.
The PIC simulations can self-consistently evolve the relativistic plasmas at their kinetic microscales.
The phase coherence of the plasma particles is achieved as the particles move collectively in self-consistently generated electrostatic waves.

In PIC simulations, the electric currents contain information about the coherence because the individual contributions of plasma particles are added up to the currents as functions of space and time.
If plasma particles oscillate collectively in a plasma region, oscillating electric currents can be formed, and coherence is achieved.
Nonetheless, if particles do not oscillate in phase, this approach leads to the mutual canceling of their currents (in an average over a few simulation time steps), and the emission is incoherent.
Because the current contributions are intrinsically added up in PIC simulations, only little post-processing is required.
In comparison with tracking individual particles in postprocessing, this approache reduces the number of simulation output data and postprocessing power (assuming that there are more particles than grid cells in the simulation).
The currents obtained from 1D PIC simulations allow calculating the emitted power as a function of the frequency and emission angle.



\subsection{Calculation steps}
Our approach for calculating LAE for the plasma instabilities has three steps.
(1) 1D PIC simulation of the instability is carried out (Appendix~\ref{methods1}). The electric currents on the grid cells are the main output for step 2 below.
The currents include aggregated particle motions as functions of space $x$ and time $t$.
The simulations were carried out in the plasma frame with the Lorentz factor $\gamma_\mathrm{s} = 100$ in the pulsar frame.
(2) Calculation of the radiation power from the electric currents in the plasma frame as a function of angle and frequency (Appendix~\ref{methods2}).
To calculate the power, we developed a novel specialized numerical model that determines LAE from space- and time-evolving electric currents from the PIC simulations.
The currents intrinsically include information about particle coherence.
(3) Relativistic beaming of the radiation power from the plasma frame to the pulsar frame (Appendix~\ref{methods3}).

\begin{figure}
        \includegraphics[width=0.49\textwidth]{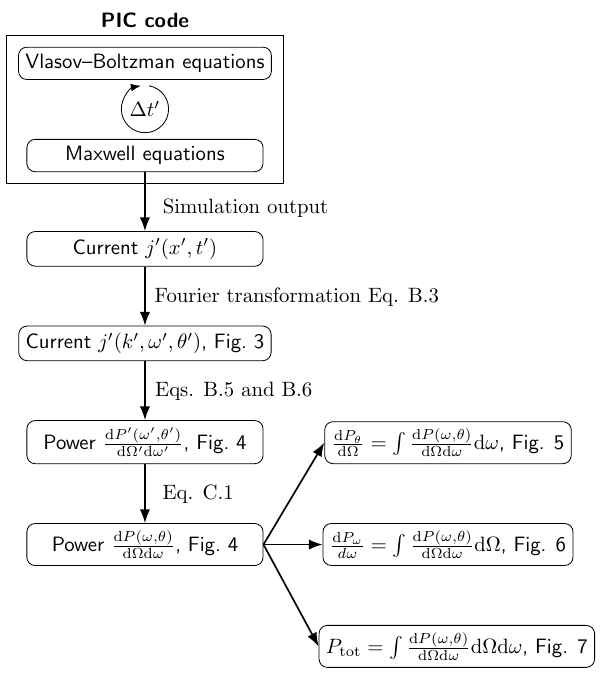}
        \caption{
        Flowchart of the calculation of LAE from the self-consistent PIC code to the radio power properties.
        The equations and figures in which the results are shown are denoted.
        The relativistic transformation of the radiation power $P$ is from the plasma frame (quantities denoted by primes) to the pulsar frame (quantities without primes). Individual symbols are described in the text.
        }
        \label{fig0}
\end{figure}

The individual steps for the calculation together with the equations and figures are summarized in Fig.~\ref{fig0} and in Appendices~\ref{methods1}, \ref{methods2}, and \ref{methods3}.
Although the variables in Appendices~\ref{methods1} and \ref{methods2} are considered in the plasma reference frame, they are not denoted by a prime for better readability of the equations.
In Appendix~\ref{methods3} the primes are used again.


\section{Results} \label{sec:results}

We estimated the LAE power of linearly accelerated particles by the streaming instability and plasma bunch interaction.
We assumed a transformation Lorentz factor of the plasma in 1D PIC simulations along the magnetic field direction as $\gamma_\mathrm{s}=100$,
which produces maxima of the power at frequencies $\sim$500~MHz in the pulsar frame.
Moreover, similar values of $\gamma_\mathrm{s}$ were found for the secondary particles created in the gap regions \citep{Arendt2002}.
As the radio emission is formed in the open magnetic field lines, we assumed that the plasma bunches and associated electrostatic waves move radially outward in the direction away from the star.

We assumed spherical coordinates of the line of sight $(r, \theta, \varphi)$, where $r$ is the distance, and $\theta$ is the polar angle, where $\theta = 0$ is the direction along the local magnetic field in which the plasma frame moves, and $\varphi$ is the azimuthal angle.
Moreover, we considered these coordinates in the plasma frame $(r', \theta', \varphi')$ and in the pulsar frame $(r, \theta, \varphi)$. 
From the 1D definition, the radiation power is symmetric in the azimuthal angle $\varphi'$ and $\varphi$.
We also note that we strictly denote all variables in the plasma frame by primes.

\begin{figure*}
        \includegraphics[width=0.49\textwidth]{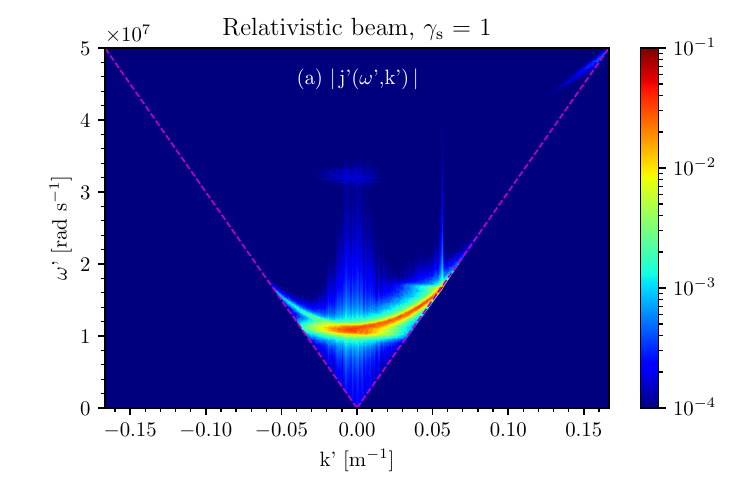}
        \includegraphics[width=0.49\textwidth]{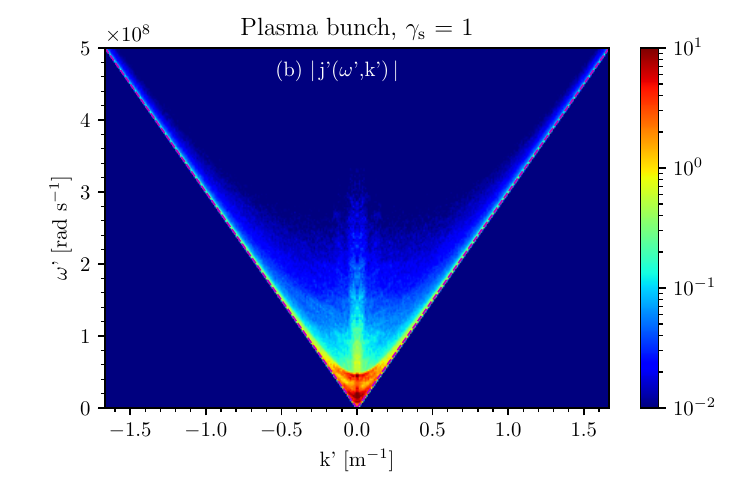}
        \caption{Electric current density as a function of frequency and wave number for relativistic streaming instability and interacting plasma bunches in the plasma reference frame.
                Both instabilities are selected in the time interval $\omega_\mathrm{p}t=0-3500$.
                Subluminal waves, which do not contribute to the emission, are set to zero.
                \textit{Dashed magenta lines:} Light lines $k = \pm \omega/c$.
                Positive wave numbers correspond to the direction away from the star.
        }
        \label{fig1}
\end{figure*}

Figure~\ref{fig1} shows the electric current in the Fourier space obtained for the relativistic streaming instability and interacting plasma bunches in their plasma frames.
Because the subluminal waves do not contribute to the emission flux as follows from Eqs.~\ref{eq:A3} and \ref{eq:dispersion}, they were set to zero.
The currents are distributed closer to $\omega'=0, k'=0$ for the plasma bunch interaction model compared to the relativistic beam instability model.
A broad range of wave frequencies is excited by the simulation of plasma bunch interaction; no specific wave mode dominates.
Generally, the emission parallel to the magnetic field direction ($\theta' \to 0$) comes from the Fourier space regions close to the light lines (dashed magenta lines),
$\lim_{\theta' \to 0} k'(\theta') = \lim_{\theta' \to 0} \frac{\omega'}{c} \cos \theta' = \frac{\omega'}{c}$ (Eq.~\ref{eq:A3}).
The emission perpendicular to the magnetic field lines comes from Fourier space regions close to $k' \approx 0$ because $\lim_{\theta' \to \frac{\pi}{2}} k(\theta') = \lim_{\theta' \to \frac{\pi}{2}} \frac{\omega'}{c} \cos \theta' = 0$.

\begin{figure*}[!ht]
        \includegraphics[width=0.49\textwidth]{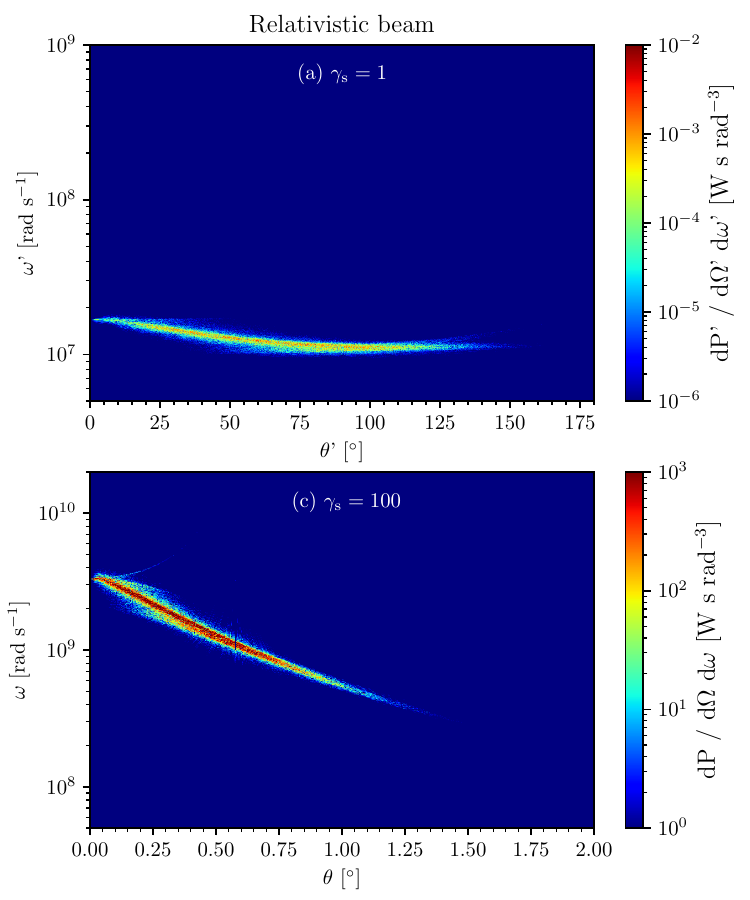}
        \includegraphics[width=0.49\textwidth]{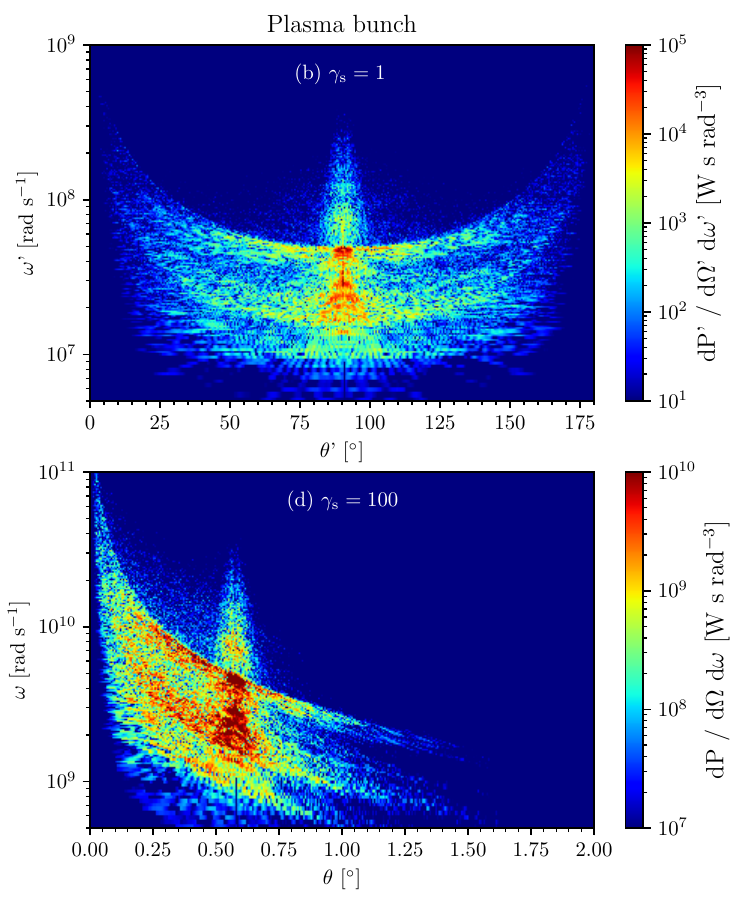}
    \caption{Average power per frequency and spatial angle units as a function of frequency and polar angle in the plasma reference frame ($\gamma_\mathrm{s}=1$, (a--b)), and in the pulsar reference frame ($\gamma_\mathrm{s}=$100, (c--d)) for the streaming instability ((a),(c)) and the interaction of plasma bunches ((b),(d)).
                The intensity and frequency scales are different.
        }
        \label{fig2}
\end{figure*}

\subsection{Average properties of the radiation power}
Figure~\ref{fig2} shows the resulting average electromagnetic power per spatial angle unit and frequency unit as a function of frequency and polar angle for the two models obtained over their whole simulation time.
The top row depicts the power in the plasma reference frame ($\gamma_\mathrm{s}=1$).
The bottom row depicts the power after its transformation into the pulsar frame ($\gamma_\mathrm{s}=100$).
In the plasma frame, the maximum power, $\mathrm{d}P'(\omega',\theta')/\mathrm{d}\Omega' \mathrm{d}\omega'$, reaches $6\times 10^{-3}$~W~s~rad$^{-3}$ for the relativistic beam and $3\times 10^{5}$~W~s~rad$^{-3}$ for the plasma bunch.
In the pulsar reference frame, the maximum powers reach $3\times10^{4}\,\mathrm{W}\,\mathrm{s}\,\mathrm{rad}^{-3}$ for the relativistic beam and $2\times10^{11}\,\mathrm{W}\,\mathrm{s}\,\mathrm{rad}^{-3}$ for the plasma bunch.
Although these maxima are reached in a broader statistically significant region of the $\omega-k$ domain for the streaming instability, they are reached only in a few points for the plasma bunch.
When these few points are eliminated, the typical highest values are smaller by approximately one order of magnitude.

The total average electromagnetic powers in the plasma frame,
\begin{equation}
        P_\mathrm{tot}=\int\iint_{\Omega}\sin(\theta) \left( \frac{\mathrm{d}P(\omega,\theta)}{\mathrm{d}\Omega \mathrm{d}\omega} \right) \mathrm{d}\theta\mathrm{d}\varphi \mathrm{d}\omega,
\end{equation}
are $1.0\times10^{8}\,\mathrm{W}$ for the relativistic beam and $8.3\times10^{15}\,\mathrm{W}$ for the plasma bunch interaction, respectively.

\begin{figure*}[!ht]
        \centering
        \includegraphics[width=0.49\textwidth]{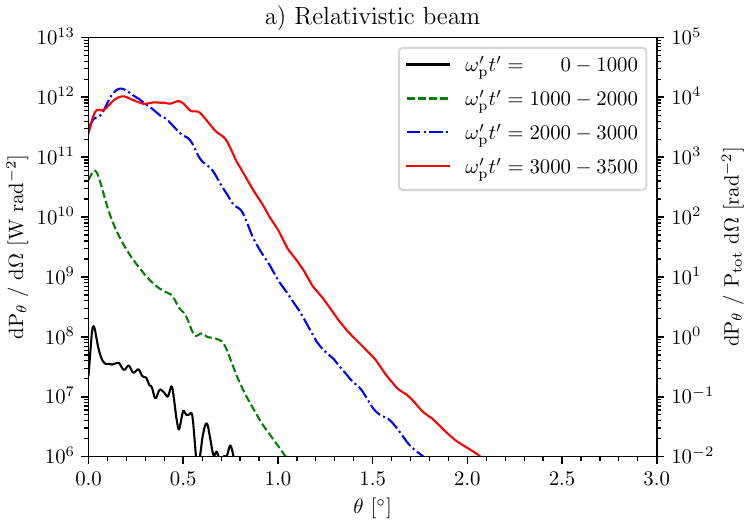}
        ~
        \includegraphics[width=0.49\textwidth]{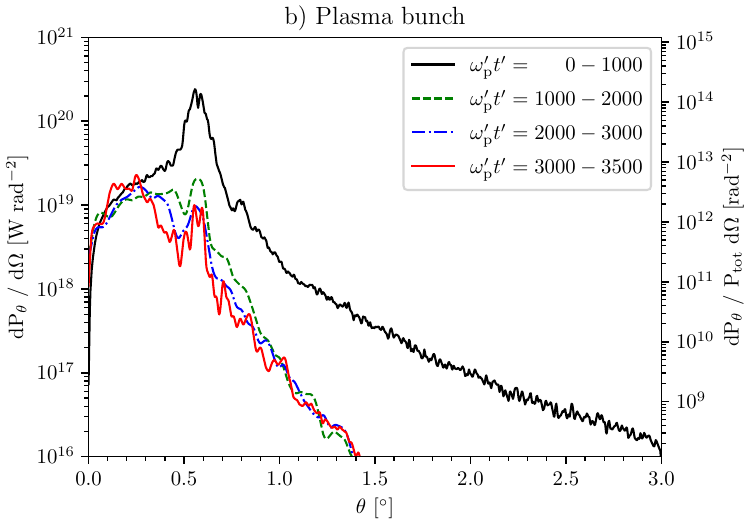}
        \caption{Average power per spatial angle unit as a function of polar angle in the pulsar frame during selected time intervals.
                The emission power (see Fig.~\ref{fig2}c--d) is integrated over all spatial angles in the pulsar reference frame.}
        \label{fig3}
\end{figure*}

\subsection{Time evolution of the radiation power}
We analyze the emission power only in the pulsar frame below.
Figure~\ref{fig3} shows the average power as a function of the polar angle.
Several time intervals are selected.
The power per spatial angle and frequency units was integrated over all frequencies,
\begin{equation}
        \frac{\mathrm{d}P_\theta}{d\Omega} = \int \left( \frac{\mathrm{d}P(\omega,\theta)}{\mathrm{d}\Omega \mathrm{d}\omega} \right) \mathrm{d}\omega,
\end{equation}
and normalized to the total power $P_\mathrm{tot}$.
For the relativistic beam, the power close to $\theta' \approx 0$ was  first enhanced at the time interval $\omega_\mathrm{p}'t'=1000-2000$.
This region corresponds to the emission of the initially most unstable superluminal waves that overlap with the light line ($\omega' = k'c$) in the $\omega'-k'$ space. \citep{Benacek2021a}.
As the superluminal wave power grows closer to $k'=0$ as well ($\theta'\rightarrow\pi/2$ in the plasma frame), the angular width of the emission increases.
In the case of the plasma bunch interaction, a power peak is reached at $\theta\approx 0.56\,^\circ$ in the time interval $\omega_\mathrm{p}'t'=0-1000$.
Then, the peak power decreases and shifts to smaller angles.
The emission angular widths become narrower and form tails from the maxima to larger angles.


\begin{figure*}[!ht]
        \includegraphics[width=0.49\textwidth]{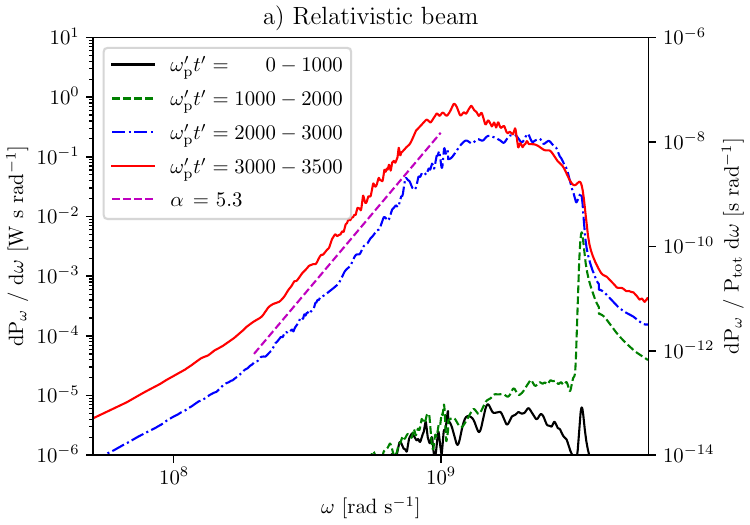}
        ~
        \includegraphics[width=0.49\textwidth]{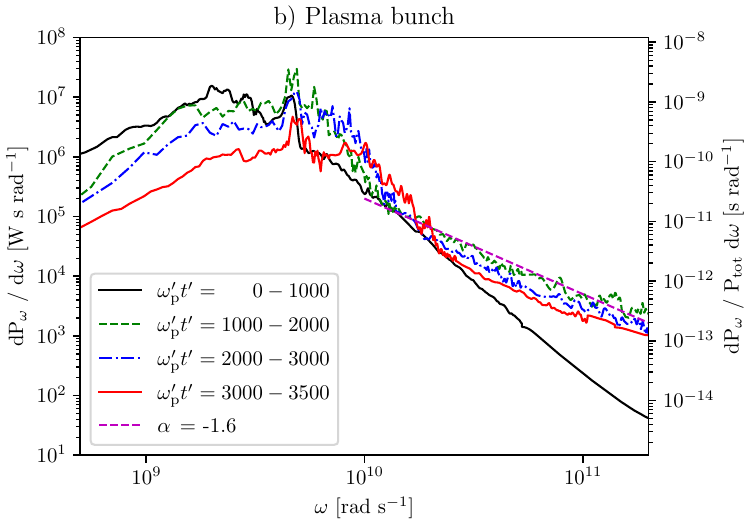}
        \caption{Average power per frequency unit as a function of frequency in the pulsar frame during selected time intervals.
                The power (see Fig.~\ref{fig2}c--d) is integrated over all spatial angles in the pulsar reference frame.
                Straight dashed lines: Power-law functions with indices $\alpha=5.3$ and $\alpha=-1.6$.}
        \label{fig4}
\end{figure*}
The average power in the pulsar reference frame is presented as a function of the frequency in Fig.~\ref{fig4}.
The power per frequency unit is the integral over all spatial angles,
\begin{equation}
        \frac{\mathrm{d}P_\omega}{d\omega} = \iint_{\Omega}\sin(\theta) \left( \frac{\mathrm{d}P(\omega,\theta)}{\mathrm{d} \Omega \mathrm{d}\omega} \right)\mathrm{d}\theta \mathrm{d}\varphi.
\end{equation}
The straight dashed lines in the Figs. correspond to power-law functions with an index $\alpha$.
For the relativistic beam (Fig.~\ref{fig4}a), the highest intensity first grows at a frequency $3.4\times10^{9}\,\mathrm{rad}\,\mathrm{s}^{-1}$ (corresponding approximately to the frequency of the unstable subluminal waves) in the time interval $\omega_\mathrm{p}'t'=1000-2000$.
Later, the power maximum is enhanced, shifting to lower frequencies and broadening.
The spectrum is characterized by a steep decrease in the wave power at a frequency $\sim$$4\times10^{9}\,\mathrm{rad}\,\mathrm{s}^{-1}$ at all times.
This decrease is also shown in Fig.~\ref{fig2}c at angles $\theta\approx0.1^{\circ}$.
We found that this frequency corresponds to the highest frequency $\omega_\mathrm{max}' > \omega_\mathrm{p}'$ of superluminal waves in the simulation.
$\omega_\mathrm{max}'$ is determined by the point in the $\omega'-k'$ domain in which the L-mode branch crosses the light line, that is, it changes from superluminal to subluminal mode \citep{Rafat2019a}.
Higher-frequency wave modes ($\omega'>\omega_\mathrm{max}'$) are subluminal, and they do not generate electromagnetic waves in the 1D limit.
At later times ($\omega_\mathrm{p}'t'=2000-3500$), the low-frequency part of the spectrum can be roughly approximated by a steep power-law function with index $\alpha\sim5.3$.

In the plasma bunch interaction, the spectrum slightly broadens in time within the frequency interval $\sim$$(3-75)\times10^{9}\,\mathrm{rad}\,\mathrm{s}^{-1}$, and it enters the higher-frequency part of the spectrum and develops power laws.
The corresponding specific power-law indices are $-3.1,-1.6,-1.7$, and $-1.7$ (for the time intervals in the order as presented in the figure).
The estimation error is $\pm0.2$.

\begin{figure*}[t]
        \includegraphics[width=0.49\textwidth]{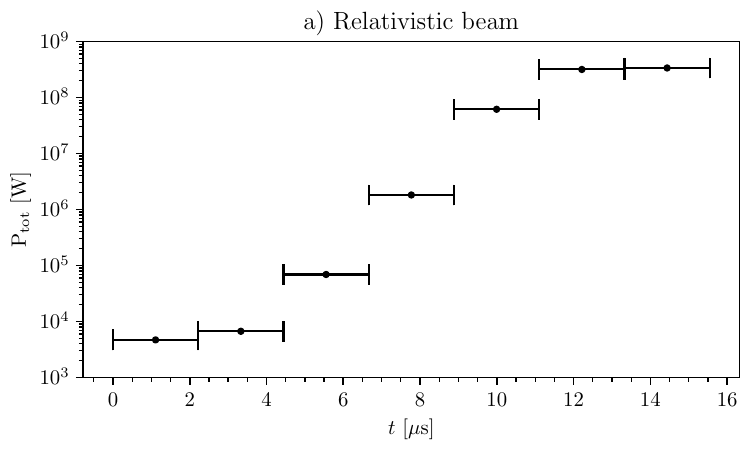}
        \includegraphics[width=0.49\textwidth]{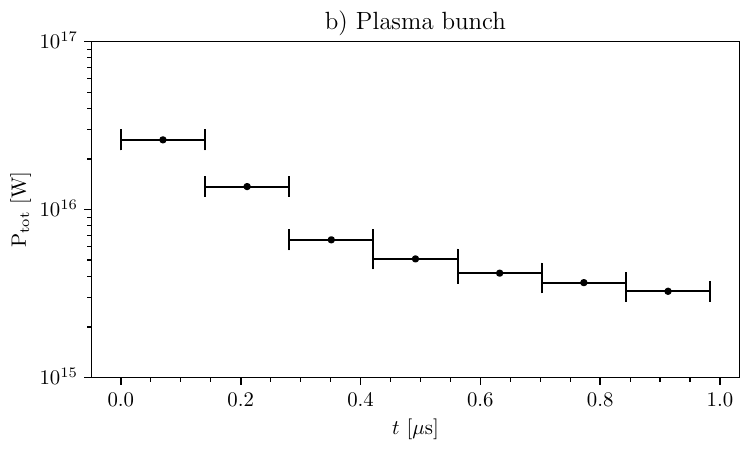}
        \caption{Evolution of the total electromagnetic power in the pulsar reference frame.
                The time intervals are denoted by horizontal bars.
                The emission power is integrated over all frequencies and spatial angles.}
        \label{fig5}
\end{figure*}
Figure~\ref{fig5} depicts the evolution of the total power, $P_\mathrm{tot}$, in the pulsar reference frame ($\gamma_\mathrm{s}=100$).
The horizontal bars correspond to the time intervals for which the data were selected in the simulation reference frame.
The initial power $\lesssim10^{4}\,\mathrm{W}$ of the relativistic beam instability (Fig.~\ref{fig5}a) is mostly due to the particle (current density) noise in the simulations.
This type of emission is incoherent, however.
Coherence is obtained in the later evolutionary stage when the coherent emission of the instability prevails.
Starting at $\omega_\mathrm{p}'t'\approx1000$, the total emission power exponentially rises 
and saturates at $3.4\times10^{8}\,\mathrm{W}$.
The plasma bunch evolution (Fig.~\ref{fig5}b) starts at a total emission power $2.6\times10^{16}\,\mathrm{W}$, as the instability develops in times $\sim\omega_\mathrm{p}'^{-1}$.
The incoherent emission by the noise is negligible in this bunch interaction case since the beginning of the simulation.
Then, its power decays and saturates at $\approx3.2\times 10^{15}\,\mathrm{W}$.

\begin{figure}[t]
        \includegraphics[width=0.49\textwidth]{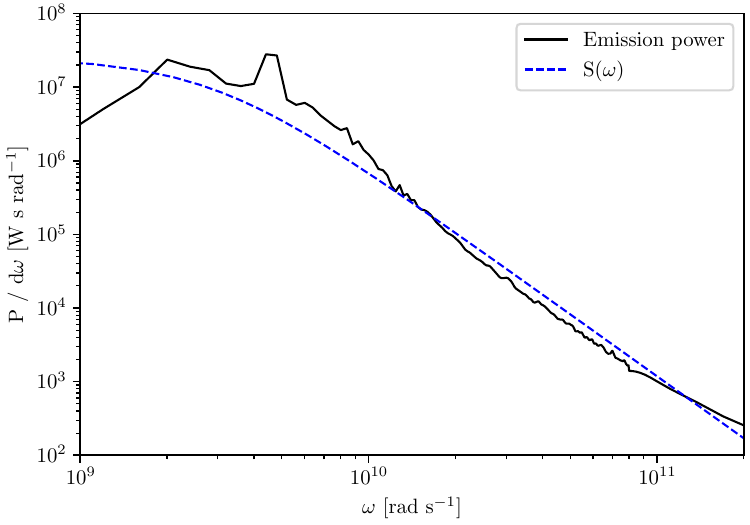}
        \caption{
                Fit of the average power per frequency unit as a function of the frequency during the whole simulation time of the interacting plasma bunches ($\gamma_\mathrm{s}=100$).
                The fit is given by Eq.~\ref{eq:Lorentz-function}.
                For the fit parameters, we refer to the text.
        }
        \label{fig9}
\end{figure}

\begin{figure*}
        \includegraphics[width=\textwidth]{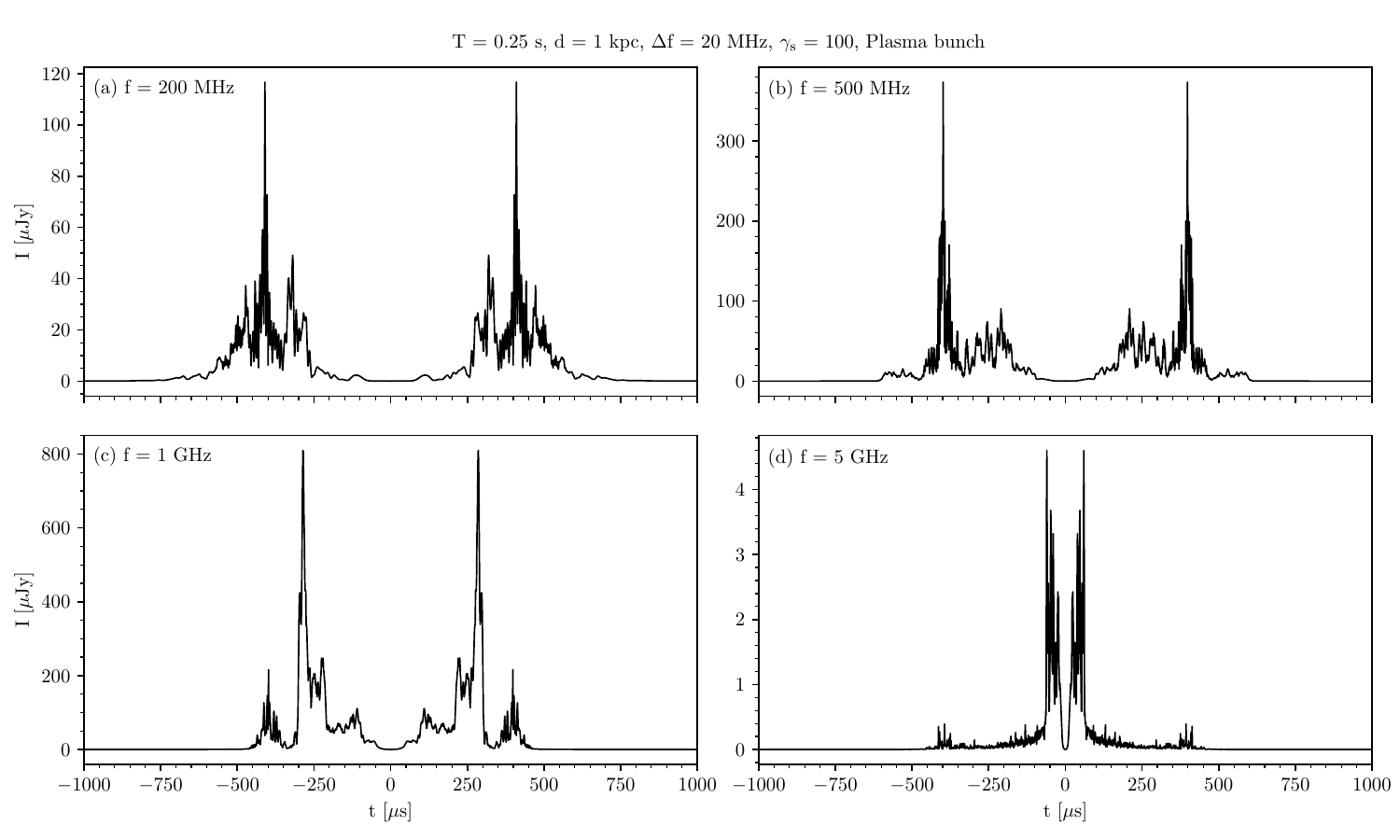}
        \caption{
                Intensities of electromagnetic waves by one plasma bunch interaction as seen from the distance $d=1\,\mathrm{kpc}$, assuming a pulsar with a rotation period $T_\mathrm{pulsar}=0.25\,\mathrm{s}$, a frequency bandwidth $\Delta{f}=20\,\mathrm{MHz}$, $\gamma_\mathrm{s}=100$, a time interval $\omega_\mathrm{p}'t'=0-3500$ in the simulation reference frame, and neglecting radiation transfer effects.
                The intensity profile can be understood as cuts for arbitrary frequency along the polar angle in the frequency--angle space (see Fig.~\ref{fig2}).
                We assume that the center of the emission cone crosses the observer at the time $t(\theta=0)=0$.
        }
        \label{fig6}
\end{figure*}

\subsection{Predictions for the observed properties}

The plasma bunch interaction emits significantly more powerful radiation than the relativistic beam instability.
Therefore, we analyze the emission properties only for the bunch interaction further in Figs.~\ref{fig9} and \ref{fig6}.

Figure~\ref{fig9} shows the average radio spectrum.
The data were taken from the whole simulation time, $\omega_\mathrm{p}'t'=0-3500$, as a function of frequency.
We fit the average spectrum by an empirical function that was used to fit the observed pulsar spectra \citep{Lohmer2008},
\begin{equation} \label{eq:Lorentz-function}
        S(\omega) = \frac{S_0}{(1 + \omega^2 \tau_\mathrm{e}^2)^{\zeta + 1}}.
\end{equation}
The obtained fit parameters are $S_0 = (2.5\pm0.3)\times 10^{7}$~W~s~rad$^{-1}$, $\tau_\mathrm{e} = (0.35\pm0.2)$~ns, and $\zeta = (0.4\pm0.05)$.

Figure~\ref{fig6} shows intensity profiles for one bunch interaction as they would be seen by an observer at a distance $d=1\,\mathrm{kpc}$ for a pulsar rotating with a period $T_\mathrm{pulsar}=2\pi/\Omega_\mathrm{c} = 0.25\,\mathrm{s}$ for four different frequencies ($200\,\mathrm{MHz}$, $500\,\mathrm{MHz}$, $1\,\mathrm{GHz}$, and $5\,\mathrm{GHz}$) at which pulsars are observed.
The frequency bandwidth is $\Delta{f}=20\,\mathrm{MHz}$.
We assumed that the observer can see the emission regions from changing angles $\theta$ as the star rotates.
We also assumed that the average power properties were constant over the whole time as the observer crosses the radio beam.
Because the rotation time is significantly longer than the simulation time, this approach is valid when one bunch interaction ends and is immediately replaced by the emission of the following interacting bunch, which has the same emission properties.
The approach is valid because the pulse crossing time is much longer than the simulation and the radio-emitting time intervals.

If at time $t=0$ the emission angle corresponds to~$0\,^\circ$, the emission is considered along the magnetic field at the moment when the center of the emission cone just crosses the observer.
The observed intensity is the average power estimated as
$I=\mathrm{d}P_\theta/(d^2\Delta{f})$.
The time is related to the emission angles as $t=\theta T_\mathrm{pulsar}/(2\pi)$, where $T_\mathrm{pulsar}=0.25\,\mathrm{s}$ is the pulsar period.
Because the emission is assumed to be symmetric for the azimuthal axis in spherical coordinates, the intensity profiles are reversible in time (for $t<0$ and for $t>0$).
We note that the numerical resolution of frequencies and wave numbers gives the shortest timescale of the intensity fluctuations.
For the real plasma bunch, the typical fluctuation timescales would be shorter.

\section{Discussion} \label{sec:discussion}

We have calculated the radio power properties of the LAE process via the coherent antenna principle in the pair plasma of neutron star magnetospheres.
We have carried out PIC simulations that described the collective and nonlinear plasma evolution at kinetic microscales.
We considered the relativistic streaming instability and plasma bunch interaction, which have been proposed as possible emission sources of pulsar coherent radio waves.
The wave power properties were obtained from the plasma currents directly resulting from the studied plasma instabilities.
The approach allowed us to estimate the coherent power due to the coherent adding up of individual contributions of plasma particles with respect to their phase.

\subsection{Emission properties of the studied instabilities}
We have found that the maximum of the estimated total power emitted by a plasma bunch interaction, $\approx$$2.6\times10^{16}\,\mathrm{W}$, exceeds the total power caused by the studied streaming instability region ($3.4\times10^{8}\,\mathrm{W}$) by eight orders of magnitude if the maxima of their spectra are at $\sim$1\,GHz.
Moreover, $\sim$$4\times (10^1-10^5)$ simultaneously emitting plasma bunches can account for the total pulsar radio emission power $10^{18}-10^{22}\,\mathrm{W}$ ($10^{25}-10^{29}\,\mathrm{erg}\,\mathrm{s}^{-1}$).
In the case of the streaming instability, at least $\sim$$3\times(10^9-10^{13})$ of these emitting regions would be necessary to account for the observed power by the streaming instability.

We estimated the number of contributing bunches assuming that their independent coherent emissions are incoherently summed to yield the total radio power of the pulsar.
Each one of the bunch emissions is still being produced by a coherent mechanism.
A similar composition of the emission model was considered for the coherent curvature emission mechanism, for instance, in which coherently radiating plasma charges are incoherently added up to obtain the total emission flux \citep{Melikidze2000}.
Moreover, because we assumed no mutual coherence between individual emitting bunches, the emission regions in the bunch trains and also the trains themselves may be distributed in a relatively large spatial region.
The region might extend to several stellar radii, for example.

Nevertheless, bunches produced by a gap region might evolve similarly in the train because the emission regions of the bunches might be close to each other.
Therefore, their emissions might also interact coherently in larger groups of bunches if the emissions of bunches are allowed to add up with respect to the wave phases. 
Then, a lower number $N \gtrsim 6-600$ of bunches would be required to explain the total observed flux.
Furthermore, to coherently add up the bunch emissions, the bunches must be located in a relatively compact region, which might be possible for this number.
We also expect that many bunch interactions in the compact region are simultaneous, although the interactions and emissions of sequentially produced bunches are shifted in time by the spark repetition interval.
However, the exact estimation of how the emissions can add up requires a study of the radiation propagation at magnetospheric distances \citep{Beskin2012}.
Moreover, as was shown by \citet{Bransgrove2022} and \citet{Cruz2021}, for example, the scenario in the polar cap region can be rather complex and requires the inclusion of propagation and superposition effects, with the LAE being the initial broadband emission process.

The total power depends on the properties of the plasma instabilities and the intensities of the electric currents induced by oscillating particles.
The particle oscillates in electrostatic superluminal L-mode waves self-consistently produced by the PIC simulations as solutions of the electrostatic permittivity tensor component $\Lambda_{33}=0$ (see \citet{Rafat2019a}).
For the streaming instability, most of the emission power depends on the formation of soliton-like superluminal waves, which are generated for inverse temperatures $\rho\geq1.66$ and Lorentz factors $\gamma_\mathrm{s}>40$ \citep{Benacek2021a}.
If these plasma properties are not fulfilled, the soliton-like superluminal waves are not produced, and their intensities, and therefore the emissions, are weaker than in the case with the soliton waves.

The emission power of the plasma bunch interaction mainly depends on the drift speeds between particle species.
Nevertheless, only for nonzero drift speeds between particle species does the energy of superluminal waves significantly exceed the energy of subluminal waves, and a significant amount of emission may be produced.
The wave power increases nonlinearly with the drift speed.
In a supplemental simulation, in which we kept the same parameters as in the bunch simulation above, but increased the drift speed from $u_\mathrm{d}'/c=10$ to $u_\mathrm{d}'/c=100$ and kept the transformation Lorentz factor $\gamma_\mathrm{s} \approx 100$, the total emission power increased from $\approx2.6\times10^{16}\,\mathrm{W}$ to $1.8\times10^{20}\,\mathrm{W}$.
We expect only a weak dependence of the superluminal wave energies on the plasma temperature.
The main reason is that the superluminal waves are generated by ambipolar diffusion at the bunch edges.
The ambipolar diffusion is mostly influenced by the drift-speed difference between the expanding particle species, while the plasma temperature $T'$ represents only a correction term to this expansion because of the relation $T' \sim \langle \gamma' \rangle \ll \gamma_\mathrm{d}' \equiv \sqrt{1 + u_\mathrm{d}'^2}$, where $\langle\rangle$ denotes an average over the particle velocity distribution in a plasma region.

A direct comparison of our results with already published calculations of pulsar LAE as a coherent radio emission mechanism is difficult.
We did not find any other approach that calculates the same properties for all plasma particles, but only for particles characterized by average plasma properties (e.g., \citet{Melrose2009a,Melrose2009b}).
As was shown by \citet{Benacek2021b}, the plasma properties significantly change in the emission region.
The generated electric fields in our simulations cannot be approximated by uniform slabs of static electric fields, as were used for the known analytical calculations of single-particle emissions.
Instead, the electric fields in the simulations oscillate over a wide range of frequencies in various positions $x$, some regions keeping a wave coherence.
Furthermore, the previous investigations were not compared with observations at all.

We can estimate over which time, $\Delta t$, the whole energy of the electrostatic superluminal L-mode waves, $E_\mathrm{tot}$, is emitted with the highest emission power, $P_\mathrm{max}$.
In the pulsar reference frame, this time is $\Delta t = E_\mathrm{tot}/P_\mathrm{max} \sim50\,\mu\mathrm{s}$ for the relativistic beam instability and $\Delta t \sim68\,$ns for the plasma bunch interaction.
For comparison, the total simulation time of the streaming instability was $15.6\,\mu\mathrm{s}$ and that of the plasma bunch interaction was $984\,n\mathrm{s}$.
Therefore, it can be expected that the emission process does not significantly influence the evolution of the relativistic beam instability.
However, most of the electrostatic energy in the plasma bunch is emitted during the simulation time.
Hence, our results must be cautiously interpreted for times $\omega_\mathrm{p}'t'\gtrsim1000$.
Moreover, the question arises whether the plasma bunch does not radiate most of its energy before another emission mechanism (e.g., the relativistic plasma emission) has time to develop before it starts to radiate.

The LAE can be considered to be similar in some aspects to the coherent curvature emission that was proposed as one of the competing coherent emission mechanisms \citep{Buschauer1976,Melikidze2000,Yang2018}.
In the case of LAE, the plasma particles oscillate along magnetic field lines undergoing acceleration in the same direction.
In the reference frame of the oscillation center, the emission of an oscillating particle has its highest intensity in the perpendicular direction to the magnetic field, and its lowest intensity is along the magnetic field.
This emission pattern is then relativistically beamed into a narrow emission cone along the magnetic field.
The O-mode polarization (electric component oscillating in the same direction as the projected magnetic field) remains the same in both relativistic frames because the relativistic transformation is in the direction of the magnetic field lines.
The coherent curvature emission differs in the direction of the acceleration and polarization.
As a plasma particle propagates at a constant velocity along a curved magnetic field, it undergoes a constant acceleration perpendicular to the magnetic field.
In the particle reference frame, the highest emission is in the plane perpendicular to the acceleration.
Similarly, the emission is relativistically beamed into a narrow cone along the magnetic field.
The polarization angle is in the direction of the particle acceleration and is seen by an observer in the direction of the magnetic field line.
Because both mechanisms have very similar radio properties due to the relativistic beaming, a differentiation of both mechanisms in observations can be difficult.

\subsection{Predictions for the observed properties}
This section describes the radio emission properties and their possible relation to observations.
We assumed that the radio waves are not significantly influenced by the plasma dispersion effects while the waves propagate through the magnetosphere toward an observer.

The emission power of both instabilities is directed into a narrow cone in the pulsar (observer) reference frame, which is mostly provided by the relativistic transformation.
In the plasma reference frame, the most intense electromagnetic waves are emitted approximately perpendicular to the magnetic field.
After the relativistic transformation into the pulsar frame, the emission narrows into an angle $\theta_0 \lesssim 1 / \gamma_\mathrm{s} \approx 0.57\,^{\circ}$.
Nonetheless, the emission width is narrower than $\theta_0$ at higher frequencies 
and wider at lower frequencies: the cone angle decreases with increasing frequency in the whole frequency interval.
For example, for the bunch interaction, the emission is narrower at $\gtrsim 6\times10^{9}\,\mathrm{rad}\,\mathrm{s}^{-1}$ and wider at frequencies $\lesssim 6\times 10^{9}\,\mathrm{rad}\,\mathrm{s}^{-1}$.

The total power depends on the relativistic factor as follows from the transformation of radiation from the plasma to the observer reference frame, $P_\mathrm{tot}\sim\gamma_s^2$, $\gamma_\mathrm{s} \gg 1$ \citep{Rybicki1986}.
The relativistic factor is typically assumed to be in the range $\gamma_\mathrm{s}=10^{2}$--$10^4$ \citep{Arendt2002}.
For example, an increase in the Lorentz factor from the lower limit of the interval $\gamma_\mathrm{s}=100$ to $\gamma_\mathrm{s}=10^4$ in the bunch interaction would increase the total emitted wave power from $2.6\times10^{16}\,\mathrm{W}$ to $2.6\times10^{20}\,\mathrm{W}$. However, as a consequence, the observed frequencies would also grow by a factor of $\sim$$100$.

The frequency range of the flat parts of the spectra of both instabilities around $10^{10}$\,rad\,s$^{-1}$ broadens in time.
However, this effect does not significantly influence the cutoff frequency for the streaming instability and power-law index of the bunch interaction.
Moreover, we analyzed an interaction of smaller bunches in a supplemental simulation and found that the frequency range of the flat spectral region increases to higher frequencies for smaller bunch sizes.
The frequency increase is caused by superluminal waves with shorter wavelengths and higher frequencies generated at plasma density gradients.

If the center of the emission cone ($\theta\approx0$) does not cross the observer, only low-frequency waves can be observed.
A similar effect, for example, is known for the main pulse of the Crab pulsar \citep{Hankins2015}.
The upper frequency limit is given by the smallest angular distance from the cone center.

The LAE polarization obtained from the 1D simulations is 100~\% linearly polarized along the magnetic field in the simulation reference frame.
As the relativistic transformation is applied along the same axis, the polarization is still in the same O-mode direction.
From the point of the observed emission cone, the polarization vector is always directed to the center of the emission cone, and it is independent of the emission frequency.
If the emission cone center crosses the observer (as the pulsar rotates), the polarization profile of the angle is a step function by an angle 180\,$^\circ$.
Otherwise, the transition is smoother.
For more detailed polarization properties, 2D or 3D simulations with the inclusion of geometrical propagation effects are necessary.

\subsection{Specific predictions for interacting plasma bunches}
As the total emission power of the plasma bunch interaction exceeds the emission of the streaming instability, we analyze its emission properties in this section.
The power-law indices of the emission spectrum are between -3.1 and -1.6 for frequencies $\gtrsim10^{10}\,\mathrm{rad}\,\mathrm{s}^{-1}$.
This power law is close, at least for time intervals $\omega_\mathrm{p}'t' \geq 1000$, to the average observed pulsar spectral index $\approx-1.4$ \citep{Bilous2016}.
Even for $\omega_\mathrm{p}'t' < 1000$, the power law is still in the observed range between -3.5 and 0.
However, the discrepancy from the observed emission spectrum might arise when the emission is generated at various plasma frequencies (plasma densities) and for various Lorentz factors $\gamma_s$ (bunch velocities in the observer frame).

The microsecond oscillations in the intensity of the waves (Fig.~\ref{fig6}) of bunch interaction might be caused by positive and negative wave interference for specific angles $\theta$.
While this emission property can support strong temporal fluctuations on kinetic timescales, the long-time average (e.g., over hours) over many plasma bunch interactions in different regions of the pulsar magnetosphere can remain stable for given pulsar observed properties (if the general magnetospheric parameters remain constant).

The width of the emission pulse of the plasma bunch interaction at high frequencies is too narrow for some real pulses.
However, the emission profile might be broadened because the pulsar pulse is formed by several simultaneously emitting bunches, assuming that individual emission regions radiate into a slightly different angle.
If the emissions from all bunches occur into an angle $\Delta \theta$ that is larger than the emission width of one bunch at high frequencies but still smaller than the emission width of one bunch at low frequencies (e.g., $\Delta \theta = 0.2\,^{\circ}$ in our case), the resulting relative emission width at high frequencies can be significantly widened to the angular width $\Delta \theta$.
In contrast, the effect on the low-frequency emission, which has a significantly broader angular width than $\Delta \theta$, will not be as strongly affected.
As a result, this effect might produce a larger relative angular widening of emission at high frequencies than is shown in Fig.~\ref{fig2}d.

There is no direct relation between the frequency of emitted electromagnetic waves from the plasma bunch interaction and the emission height in the magnetosphere as the interacting plasma bunches (small in comparison with the plasma density scale height) simultaneously radiate in a wide range of frequencies.
The plasma bunch mostly radiates at its density gradient.
The local current oscillation frequencies (emission frequencies) are frequencies of the superluminal electrostatic waves.
The oscillation frequency may be characterized for $k'=0$ as $\omega' \approx \omega_0' = \langle \gamma'^{-3} \rangle \omega_\mathrm{p}'(x')$ \citep{Rafat2019a}, where $\omega_\mathrm{p}'(x')$ is the local plasma frequency, which varies along the bunch and $\langle \rangle$ denotes the average over the local particle velocity distribution function, $\omega_0' < \omega_\mathrm{p}'(x')$ for $\rho' \sim 1$.

Because the plasma bunches radiate approximately simultaneously at all frequencies in a small region (in comparison with the scales of the magnetosphere), the relation of radius--frequency mapping is invalid.
This conclusion was observationally supported by \citet{Hassall2012} and \cite{Hassall2013}, who also indicated that the emission is instantaneous in a broad range of frequencies and that the emission occurs without any light travel-time delays caused by a significant emission radius to frequency mapping for various frequencies.

\section{Conclusions} \label{sec:conclusion}
By taking the collective and nonlinear plasma evolution into account, we have found that the coherent LAE by the antenna principle is a promising emission mechanism of interacting plasma bunches.
The mechanism has several of the observed features of pulsar radio signals.
Hence, the LAE mechanism should not be neglected in future considerations of pulsars and fast radio bursts.
Furthermore, in order to obtain more precise emission estimates,
2D or 3D fully electromagnetic PIC simulations should be carried out.
Consideration of transverse waves, their coherence, absorption, and propagation effects will provide even better insight into these processes.

\begin{acknowledgements}
We thank the anonymous referee for helpful comments that improved the quality of the manuscript.
The authors are grateful to Kuo Liu for his careful reading of the manuscript and helpful comments.
They acknowledge the support by the German Science Foundation (DFG) projects BU~777-17-1 and BE~7886/2-1. 
We acknowledge the developers of the ACRONYM code (Verein zur F\"orderung kinetischer Plasmasimulationen e.V.).
The authors gratefully acknowledge the Gauss Centre for Supercomputing e.V. (\url{www.gauss-centre.eu}) for partially funding this project by providing computing time on the GCS Supercomputer SuperMUC-NG at Leibniz Supercomputing Centre (www.lrz.de), projects pr74vi and pn73ne.
\end{acknowledgements}

%
%

\bibliographystyle{aa}
\bibliography{references}

\begin{thebibliography}{84}
\expandafter\ifx\csname natexlab\endcsname\relax\def\natexlab#1{#1}\fi

\bibitem[{{Akhiezer} {et~al.}(1975){Akhiezer}, {Akhiezer}, {Polovin},
  {Sitenko}, \& {Stepanov}}]{Akhiezer1975}
{Akhiezer}, A.~I., {Akhiezer}, I.~A., {Polovin}, R.~V., {Sitenko}, A.~G., \&
  {Stepanov}, K.~N. 1975, Oxford Pergamon Press International Series on Natural
  Philosophy, 1

\bibitem[{Arendt \& Eilek(2002)}]{Arendt2002}
Arendt, P.~N. \& Eilek, J.~A. 2002, \apj, 581, 451

\bibitem[{Arons \& Barnard(1986)}]{Arons1986}
Arons, J. \& Barnard, J.~J. 1986, \apj, 302, 120

\bibitem[{{Ben{\'a}{\v{c}}ek} {et~al.}(2021{\natexlab{a}}){Ben{\'a}{\v{c}}ek},
  {Mu{\~n}oz}, \& {B{\"u}chner}}]{Benacek2021b}
{Ben{\'a}{\v{c}}ek}, J., {Mu{\~n}oz}, P.~A., \& {B{\"u}chner}, J.
  2021{\natexlab{a}}, \apj, 923, 99

\bibitem[{{Ben{\'a}{\v{c}}ek} {et~al.}(2021{\natexlab{b}}){Ben{\'a}{\v{c}}ek},
  {Mu{\~n}oz}, {Manthei}, \& {B{\"u}chner}}]{Benacek2021a}
{Ben{\'a}{\v{c}}ek}, J., {Mu{\~n}oz}, P.~A., {Manthei}, A.~C., \&
  {B{\"u}chner}, J. 2021{\natexlab{b}}, \apj, 915, 127

\bibitem[{Benford \& Buschauer(1977)}]{Benford1977}
Benford, G. \& Buschauer, R. 1977, Monthly Notices of the Royal Astronomical
  Society, 179, 189

\bibitem[{Beskin(2018)}]{Beskin2018}
Beskin, V.~S. 2018, Uspekhi Fiz. Nauk, 188, 377

\bibitem[{Beskin {et~al.}(1993)Beskin, Gurevich, \& Istomin}]{Beskin1993}
Beskin, V.~S., Gurevich, S.~V., \& Istomin, Y.~N. 1993, {Physics of the pulsar
  magnetosphere} (Cambridge University Press)

\bibitem[{Beskin \& Philippov(2012)}]{Beskin2012}
Beskin, V.~S. \& Philippov, A.~A. 2012, \mnras, 425, 814–840

\bibitem[{Bilous {et~al.}(2016)Bilous, Kondratiev, Kramer, Keane, Hessels,
  Stappers, Malofeev, Sobey, Breton, Cooper, \& et~al.}]{Bilous2016}
Bilous, A.~V., Kondratiev, V.~I., Kramer, M., {et~al.} 2016, \aap, 591, A134

\bibitem[{Boris(1970)}]{Boris1970}
Boris, J.~P. 1970, in Proceedings of the Fourth Conference on the Numerical
  Simulation of Plasmas, Washington DC, ed. {J. Boris} (Naval Research
  Laboratory), 3--67

\bibitem[{{Bransgrove} {et~al.}(2022){Bransgrove}, {Beloborodov}, \&
  {Levin}}]{Bransgrove2022}
{Bransgrove}, A., {Beloborodov}, A.~M., \& {Levin}, Y. 2022, arXiv e-prints,
  arXiv:2209.11362

\bibitem[{{Buschauer} \& {Benford}(1976)}]{Buschauer1976}
{Buschauer}, R. \& {Benford}, G. 1976, \mnras, 177, 109

\bibitem[{Buschauer \& Benford(1977)}]{Buschauer1977}
Buschauer, R. \& Benford, G. 1977, \mnras, 179, 99

\bibitem[{Cerutti {et~al.}(2015)Cerutti, Philippov, Parfrey, \&
  Spitkovsky}]{Cerutti2015}
Cerutti, B., Philippov, A., Parfrey, K., \& Spitkovsky, A. 2015, Monthly
  Notices of the Royal Astronomical Society, 448, 606–619

\bibitem[{Chen \& Beloborodov(2014)}]{Chen2014}
Chen, A.~Y. \& Beloborodov, A.~M. 2014, The Astrophysical Journal, 795, L22

\bibitem[{Cheng \& Ruderman(1977{\natexlab{a}})}]{Cheng1977b}
Cheng, A.~F. \& Ruderman, M.~A. 1977{\natexlab{a}}, \apj, 212, 800

\bibitem[{Cheng \& Ruderman(1977{\natexlab{b}})}]{Cheng1977a}
Cheng, A.~F. \& Ruderman, M.~A. 1977{\natexlab{b}}, \apj, 214, 598

\bibitem[{{Cocke}(1973)}]{Cocke1973}
{Cocke}, W.~J. 1973, \apj, 184, 291

\bibitem[{Cruz {et~al.}(2021)Cruz, Grismayer, Chen, Spitkovsky, \&
  Silva}]{Cruz2021}
Cruz, F., Grismayer, T., Chen, A.~Y., Spitkovsky, A., \& Silva, L.~O. 2021,
  \apjl, 919, L4

\bibitem[{{Cruz} {et~al.}(2021){Cruz}, {Grismayer}, \& {Silva}}]{Cruz2020}
{Cruz}, F., {Grismayer}, T., \& {Silva}, L.~O. 2021, \apj, 908, 149

\bibitem[{{Decker}(1995)}]{Decker1995}
{Decker}, F.~J. 1995, in American Institute of Physics Conference Series, Vol.
  333, The 6th Workshop on beam Instrumentation, 550--556

\bibitem[{Eilek \& Hankins(2016)}]{Eilek2016}
Eilek, J. \& Hankins, T. 2016, J. Plasma Phys., 82, 635820302

\bibitem[{Esirkepov(2001)}]{Esirkepov2001}
Esirkepov, T. 2001, Computer Physics Communications, 135, 144

\bibitem[{Geng {et~al.}(2021)Geng, Meng, Yan, Zhang, \& Zhao}]{Geng2021}
Geng, H., Meng, C., Yan, F., Zhang, Y., \& Zhao, Y. 2021, in Proc. IPAC'21,
  International Particle Accelerator Conference No.~12 (JACoW Publishing,
  Geneva, Switzerland), 3759--3761,
  https://doi.org/10.18429/JACoW-IPAC2021-THPAB003

\bibitem[{{Ginzburg} \& {Zhelezniakov}(1975)}]{Ginzburg1975}
{Ginzburg}, V.~L. \& {Zhelezniakov}, V.~V. 1975, \araa, 13, 511

\bibitem[{{Goldreich} \& {Julian}(1969)}]{Goldreich1969}
{Goldreich}, P. \& {Julian}, W.~H. 1969, \apj, 157, 869

\bibitem[{{Griffiths}(2017)}]{Griffiths2017}
{Griffiths}, D.~J. 2017, {Introduction to Electrodynamics}, 4th edn. (Cambridge
  University Press), 620

\bibitem[{Hankins {et~al.}(2015)Hankins, Jones, \& Eilek}]{Hankins2015}
Hankins, T.~H., Jones, G., \& Eilek, J.~A. 2015, The Astrophysical Journal,
  802, 130

\bibitem[{{Hassall} {et~al.}(2012){Hassall}, {Stappers}, {Hessels}, {Kramer},
  {Alexov}, {Anderson}, {Coenen}, {Karastergiou}, {Keane}, {Kondratiev},
  {Lazaridis}, {van Leeuwen}, {Noutsos}, {Serylak}, {Sobey}, {Verbiest},
  {Weltevrede}, {Zagkouris}, {Fender}, {Wijers}, {B{\"a}hren}, {Bell},
  {Broderick}, {Corbel}, {Daw}, {Dhillon}, {Eisl{\"o}ffel}, {Falcke},
  {Grie{\ss}meier}, {Jonker}, {Law}, {Markoff}, {Miller-Jones}, {Osten}, {Rol},
  {Scaife}, {Scheers}, {Schellart}, {Spreeuw}, {Swinbank}, {ter Veen}, {Wise},
  {Wijnands}, {Wucknitz}, {Zarka}, {Asgekar}, {Bell}, {Bentum}, {Bernardi},
  {Best}, {Bonafede}, {Boonstra}, {Brentjens}, {Brouw}, {Br{\"u}ggen},
  {Butcher}, {Ciardi}, {Garrett}, {Gerbers}, {Gunst}, {van Haarlem}, {Heald},
  {Hoeft}, {Holties}, {de Jong}, {Koopmans}, {Kuniyoshi}, {Kuper}, {Loose},
  {Maat}, {Masters}, {McKean}, {Meulman}, {Mevius}, {Munk}, {Noordam},
  {Orr{\'u}}, {Paas}, {Pandey-Pommier}, {Pandey}, {Pizzo}, {Polatidis},
  {Reich}, {R{\"o}ttgering}, {Sluman}, {Steinmetz}, {Sterks}, {Tagger}, {Tang},
  {Tasse}, {Vermeulen}, {van Weeren}, {Wijnholds}, \&
  {Yatawatta}}]{Hassall2012}
{Hassall}, T.~E., {Stappers}, B.~W., {Hessels}, J.~W.~T., {et~al.} 2012, \aap,
  543, A66

\bibitem[{{Hassall} {et~al.}(2013){Hassall}, {Stappers}, {Weltevrede},
  {Hessels}, {Alexov}, {Coenen}, {Karastergiou}, {Kramer}, {Keane},
  {Kondratiev}, {van Leeuwen}, {Noutsos}, {Pilia}, {Serylak}, {Sobey},
  {Zagkouris}, {Fender}, {Bell}, {Broderick}, {Eisl{\"o}ffel}, {Falcke},
  {Grie{\ss}meier}, {Kuniyoshi}, {Miller-Jones}, {Wise}, {Wucknitz}, {Zarka},
  {Asgekar}, {Batejat}, {Bentum}, {Bernardi}, {Best}, {Bonafede}, {Breitling},
  {Br{\"u}ggen}, {Butcher}, {Ciardi}, {de Gasperin}, {de Reijer}, {Duscha},
  {Fallows}, {Ferrari}, {Frieswijk}, {Garrett}, {Gunst}, {Heald}, {Hoeft},
  {Juette}, {Maat}, {McKean}, {Norden}, {Pandey-Pommier}, {Pizzo}, {Polatidis},
  {Reich}, {R{\"o}ttgering}, {Sluman}, {Tang}, {Tasse}, {Vermeulen}, {van
  Weeren}, {Wijnholds}, \& {Yatawatta}}]{Hassall2013}
{Hassall}, T.~E., {Stappers}, B.~W., {Weltevrede}, P., {et~al.} 2013, \aap,
  552, A61

\bibitem[{{Jackson}(1998)}]{Jackson1998}
{Jackson}, J.~D. 1998, {Classical Electrodynamics, 3rd Edition}

\bibitem[{J{\"{u}}ttner(1911)}]{Juttner1911}
J{\"{u}}ttner, F. 1911, Ann. Phys., 339, 856

\bibitem[{K{\"{a}}rkk{\"{a}}inen \& Gjonaj(2006)}]{Karkkainen2006}
K{\"{a}}rkk{\"{a}}inen, M. \& Gjonaj, E. 2006, Proc. International
  Computational Accelerator Physics Conference, 35

\bibitem[{Kilian {et~al.}(2012)Kilian, Burkart, \& Spanier}]{Kilian2012}
Kilian, P., Burkart, T., \& Spanier, F. 2012, in High Perform. Comput. Sci.
  Eng. '11, ed. W.~E. Nagel, D.~B. Kr{\"{o}}ner, \& M.~M. Resch (Berlin,
  Heidelberg: Springer Berlin Heidelberg), 5--13

\bibitem[{{Kroll} \& {McMullin}(1979)}]{Kroll1979}
{Kroll}, N.~M. \& {McMullin}, W.~A. 1979, \apj, 231, 425

\bibitem[{Levinson \& Cerutti(2018)}]{Levinson2018}
Levinson, A. \& Cerutti, B. 2018, \aap, 616, A184, 35 citations (Crossref)
  [2023-02-08]

\bibitem[{{Levinson} {et~al.}(2005){Levinson}, {Melrose}, {Judge}, \&
  {Luo}}]{Levinson2005}
{Levinson}, A., {Melrose}, D., {Judge}, A., \& {Luo}, Q. 2005, \apj, 631, 456

\bibitem[{Liang {et~al.}(2022)Liang, Xia, Pukhov, \& Farmer}]{Liang2022}
Liang, L., Xia, G., Pukhov, A., \& Farmer, J.~P. 2022

\bibitem[{{L{\"o}hmer} {et~al.}(2008){L{\"o}hmer}, {Jessner}, {Kramer},
  {Wielebinski}, \& {Maron}}]{Lohmer2008}
{L{\"o}hmer}, O., {Jessner}, A., {Kramer}, M., {Wielebinski}, R., \& {Maron},
  O. 2008, \aap, 480, 623

\bibitem[{{Lu} \& {Kumar}(2018)}]{Lu2018}
{Lu}, W. \& {Kumar}, P. 2018, \mnras, 477, 2470

\bibitem[{{Lu} {et~al.}(2020){Lu}, {Kilian}, {Guo}, {Li}, \&
  {Liang}}]{LuKilian2020}
{Lu}, Y., {Kilian}, P., {Guo}, F., {Li}, H., \& {Liang}, E. 2020, Journal of
  Computational Physics, 413, 109388

\bibitem[{{Luo} \& {Melrose}(2008)}]{Luo2008}
{Luo}, Q. \& {Melrose}, D. 2008, \mnras, 387, 1291

\bibitem[{{Lyubarskii} \& {Petrova}(1998)}]{Liubarskii1998}
{Lyubarskii}, Y.~E. \& {Petrova}, S.~A. 1998, \aap, 333, 181

\bibitem[{{Manthei} {et~al.}(2021){Manthei}, {Ben{\'a}{\v{c}}ek}, {Mu{\~n}oz},
  \& {B{\"u}chner}}]{Manthei2021}
{Manthei}, A.~C., {Ben{\'a}{\v{c}}ek}, J., {Mu{\~n}oz}, P.~A., \&
  {B{\"u}chner}, J. 2021, \aap, 649, A145

\bibitem[{{Melikidze} {et~al.}(2000){Melikidze}, {Gil}, \&
  {Pataraya}}]{Melikidze2000}
{Melikidze}, G.~I., {Gil}, J.~A., \& {Pataraya}, A.~D. 2000, \apj, 544, 1081

\bibitem[{{Melrose}(1978)}]{Melrose1978}
{Melrose}, D.~B. 1978, \apj, 225, 557

\bibitem[{Melrose(1986)}]{Melrose1986}
Melrose, D.~B. 1986, {Instabilities in Space and Laboratory Plasmas}
  (Cambridge: Cambridge University Press)

\bibitem[{Melrose(2017)}]{Melrose2017b}
Melrose, D.~B. 2017, Reviews of Modern Plasma Physics, 1, 5

\bibitem[{Melrose \& Gedalin(1999)}]{Melrose1999}
Melrose, D.~B. \& Gedalin, M.~E. 1999, \apj, 521, 351

\bibitem[{{Melrose} \& {Luo}(2009)}]{Melrose2009b}
{Melrose}, D.~B. \& {Luo}, Q. 2009, \apj, 698, 124

\bibitem[{{Melrose} \& {McPhedran}(1991)}]{Melrose1991}
{Melrose}, D.~B. \& {McPhedran}, R.~C. 1991, {Electromagnetic Processes in
  Dispersive Media} (Cambridge University Press), 432

\bibitem[{{Melrose} {et~al.}(2009){Melrose}, {Rafat}, \& {Luo}}]{Melrose2009a}
{Melrose}, D.~B., {Rafat}, M.~Z., \& {Luo}, Q. 2009, \apj, 698, 115

\bibitem[{Melrose {et~al.}(2020)Melrose, Rafat, \& Mastrano}]{Melrose2020a}
Melrose, D.~B., Rafat, M.~Z., \& Mastrano, A. 2020, \mnras, 500, 4530

\bibitem[{{Michel}(2004)}]{Michel2004}
{Michel}, F.~C. 2004, Advances in Space Research, 33, 542

\bibitem[{{Nishikawa} {et~al.}(2021){Nishikawa}, {Duţan}, {K{\"o}hn}, \&
  {Mizuno}}]{Nishikawa2021}
{Nishikawa}, K., {Duţan}, I., {K{\"o}hn}, C., \& {Mizuno}, Y. 2021, Living
  Reviews in Computational Astrophysics, 7, 1

\bibitem[{Papadopoulou {et~al.}(2020)Papadopoulou, Antoniou, Argyropoulos,
  Hostettler, Papaphilippou, \& Trad}]{Papadopoulos2020}
Papadopoulou, S., Antoniou, F., Argyropoulos, T., {et~al.} 2020, Phys. Rev.
  Accel. Beams, 23, 101004

\bibitem[{Petrova(2002)}]{Petrova2002}
Petrova, S.~A. 2002, \aap, 383, 1067–1075

\bibitem[{Petrova(2013)}]{Petrova2013}
Petrova, S.~A. 2013, \apj, 764, 129

\bibitem[{{Philippov} \& {Kramer}(2022)}]{Philippov2022}
{Philippov}, A. \& {Kramer}, M. 2022, \araa, 60, 495

\bibitem[{{Philippov} {et~al.}(2020){Philippov}, {Timokhin}, \&
  {Spitkovsky}}]{Philippov2020}
{Philippov}, A., {Timokhin}, A., \& {Spitkovsky}, A. 2020, \prl, 124, 245101

\bibitem[{{Philippov} {et~al.}(2015){Philippov}, {Cerutti}, {Tchekhovskoy}, \&
  {Spitkovsky}}]{Philippov2015}
{Philippov}, A.~A., {Cerutti}, B., {Tchekhovskoy}, A., \& {Spitkovsky}, A.
  2015, \apjl, 815, L19

\bibitem[{Rafat {et~al.}(2019)Rafat, Melrose, \& Mastrano}]{Rafat2019a}
Rafat, M.~Z., Melrose, D.~B., \& Mastrano, A. 2019, J. Plasma Phys., 85,
  905850305

\bibitem[{{Rahaman} {et~al.}(2020){Rahaman}, {Mitra}, \&
  {Melikidze}}]{Rahaman2020}
{Rahaman}, S.~M., {Mitra}, D., \& {Melikidze}, G.~I. 2020, \mnras, 497, 3953

\bibitem[{{Reville} \& {Kirk}(2010)}]{Reville2010}
{Reville}, B. \& {Kirk}, J.~G. 2010, \apj, 715, 186

\bibitem[{{Rowe}(1992{\natexlab{a}})}]{Rowe1992a}
{Rowe}, E.~T. 1992{\natexlab{a}}, Australian Journal of Physics, 45, 1

\bibitem[{{Rowe}(1992{\natexlab{b}})}]{Rowe1992b}
{Rowe}, E.~T. 1992{\natexlab{b}}, Australian Journal of Physics, 45, 21

\bibitem[{Ruderman \& Sutherland(1975)}]{Ruderman1975}
Ruderman, M.~A. \& Sutherland, P.~G. 1975, \apj, 196, 51

\bibitem[{{Rybicki} \& {Lightman}(1986)}]{Rybicki1986}
{Rybicki}, G.~B. \& {Lightman}, A.~P. 1986, {Radiative Processes in
  Astrophysics} (John Wiley \& Sons, Ltd)

\bibitem[{Rylov(1978)}]{Rylov1978}
Rylov, Y.~A. 1978, Astrophysics and Space Science, 53, 377–402

\bibitem[{Sturrock(1971)}]{Sturrock1971}
Sturrock, P.~A. 1971, \apj, 164, 529

\bibitem[{{Timokhin}(2010)}]{Timokhin2010}
{Timokhin}, A.~N. 2010, \mnras, 408, 2092

\bibitem[{{Timokhin} \& {Arons}(2013)}]{Timokhin2013}
{Timokhin}, A.~N. \& {Arons}, J. 2013, \mnras, 429, 20

\bibitem[{Timokhin \& Harding(2019)}]{Timokhin2019}
Timokhin, A.~N. \& Harding, A.~K. 2019, The Astrophysical Journal, 871, 12

\bibitem[{{Urpin}(2014)}]{Urpin2014}
{Urpin}, V. 2014, \aap, 563, A29

\bibitem[{Ursov \& Usov(1988)}]{Ursov1988}
Ursov, V. \& Usov, V. 1988, \aaps, 140, 325

\bibitem[{Usov(1987)}]{Usov1987}
Usov, V.~V. 1987, \apj, 320, 333

\bibitem[{{Usov}(2002)}]{Usov2002}
{Usov}, V.~V. 2002, in Neutron Stars, Pulsars, and Supernova Remnants, ed.
  W.~{Becker}, H.~{Lesch}, \& J.~{Tr{\"u}mper}, 240

\bibitem[{Weatherall(1994)}]{Weatherall1994}
Weatherall, J.~C. 1994, \apj, 428, 261

\bibitem[{{Weatherall}(1997)}]{Weatherall1997}
{Weatherall}, J.~C. 1997, \apj, 483, 402

\bibitem[{Yang \& Zhang(2018)}]{Yang2018}
Yang, Y.-P. \& Zhang, B. 2018, The Astrophysical Journal, 868, 31

\bibitem[{{Yang} {et~al.}(2020){Yang}, {Zhu}, {Zhang}, \& {Wu}}]{Yang2020}
{Yang}, Y.-P., {Zhu}, J.-P., {Zhang}, B., \& {Wu}, X.-F. 2020, \apjl, 901, L13

\bibitem[{Yee(1966)}]{Yee1966}
Yee, K.~S. 1966, IEEE Trans. Antennas Propag., 14, 302

\bibitem[{{Zhang}(2020)}]{Zhang2020}
{Zhang}, B. 2020, \nat, 587, 45

\end{thebibliography}

\begin{appendix}

\section{Particle-in-cell simulations} \label{methods1}
Although the variables in this and the following Appendices~\ref{methods1} and \ref{methods2} are considered in the plasma reference frame, they are not denoted by a prime because all of them are in the plasma reference frame.
In Appendix~\ref{methods3} on, the primes are used again to denote the plasma frame.

The calculation of the LAE (Appendix~\ref{methods2}) requires knowledge of space-time dependent values of the electric current density in the pulsar plasma.
To obtain these quantities, we carried out simulations using the fully kinetic, electromagnetic, and explicit 1D3V version of the particle-in-cell (PIC) code ACRONYM\footnote{\url{http://plasma.nerd2nerd.org}} \citep{Kilian2012}.
The code uses a Yee lattice \citep{Yee1966}, a standard relativistic Boris particle push \citep{Boris1970}, \citet{Esirkepov2001} deposition scheme, and a recently developed and published weighting-with-time-dependence (WT4) fourth-order particle shape function \citep{LuKilian2020}.
This shape function significantly decreases the numerical noise produced by relativistically moving particles in the form of the numerical Cherenkov radiation.
The Cole-K\"ark\"ainen CK5 solver \citep{Karkkainen2006} enhances the correct wave propagation
at phase speeds close to the speed of light.
Periodic boundary conditions and current smoothing were applied.

Because any perpendicular kinetic energy of a particle should immediately be emitted by high-frequency synchrotron radiation in a time interval shorter than one simulation time step, the particles in the simulation were confined to move only along a magnetic field.
Hence, the produced electric fields are in the direction of the magnetic field.
Specifically, these electric fields are L-mode electrostatic waves formed as solutions of the electrostatic permittivity tensor component $\Lambda_{33}=0$ \citep{Rafat2019a}.
The initial electric fields in each simulation were zero.

We used the SI system of units throughout.
The Gaussian CGS system of units (for the current density) used by ACRONYM was converted.
Table~\ref{tab1} shows a summary of the initial parameters of our simulations.
If not explicitly mentioned, we used the data throughout the whole simulation box in the whole simulation interval.

\subsection{Common simulation setup}
We ran two simulations with similar initial conditions as \citet{Benacek2021b,Benacek2021a} for \textit{the relativistic streaming instability} and \textit{the interaction of plasma bunches} with an initial drift speed between electrons and positrons.
The simulation axis $x$ is located along a magnetic field line, neglecting its spatial curvature.
We used a time step $\omega_\mathrm{p} \Delta t = 0.025$ and a normalized grid cell size $\Delta x/d_\mathrm{e} = 0.05$, where $\omega_\mathrm{p}$ is the plasma frequency and $d_\mathrm{e} = c/\omega_\mathrm{p}$ is the plasma skin depth.
The current density was stored every tenth time step.
The highest frequency resolution in the plasma frame was $4\omega_\mathrm{p}$.
The simulations initially had the same number of electrons and positrons in each grid cell.

The net current was subtracted at each time step.
In addition, there is no net charge in the simulations because all species have the same number of particles per cell.
The simulation setup is allowed under the assumption of much higher densities than the average magnetospheric charge and current densities.
Although various scenarios of fluctuating magnetospheric currents can occur \citep{Levinson2005,Timokhin2010,Timokhin2013}, these effects are suppressed outside of the gap environment where the bunches may be emitting.
We consider the plasma evolution where the simulation domain is located in a plasma region in which the magnetospheric currents are compensated by a background plasma or are small enough not to influence the instability evolution.

We selected plasma temperatures and drift speeds that were shown to produce electrostatic waves with the largest amplitudes for the realistic plasma parameters from previous estimations \citep{Arendt2002,Usov2002}. 
The plasma properties we used are close to the values that were found.

The plasma frequencies were selected for both cases such that the center of the emission power spectrum in the pulsar reference frame was located close to the frequency $f_\mathrm{obs} \sim 500$\,MHz.
This frequency corresponds to typically observed pulsar radiation frequencies, for example, \citet{Bilous2016}.
In the plasma reference frame, we chose the plasma frequency $\omega_\mathrm{p} = \sqrt{2}\times 10^{7}$~rad~s$^{-1}$ for the relativistic beam simulation      and $\omega_\mathrm{p} = \sqrt{5} \times 10^{8}$~rad~s$^{-1}$ for the simulation of the bunch interaction.
Although the setup plasma frequency for the bunch interaction is higher than for the streaming instability, the plasma frequency corresponds to the center of the plasma bunch where the plasma density is the highest.
However, at the bunch edges, where most of the emission occurs, the plasma frequency is lower, comparable with the plasma frequency we used for the streaming instability.
In both cases, the emission frequencies were relativistically transformed from the plasma to pulsar reference frames.
For the $\gamma_\mathrm{s} = 100$ of secondary particles we used and assuming emission at the plasma frequency, the relativistic transformation of the emission frequency into the pulsar frame can be approximately expressed as $\omega_\mathrm{obs} \approx 2 \gamma_\mathrm{s} \omega_\mathrm{p} \approx 2\sqrt{2}\times 10^9$\,rad\,s$^{-1}$ $\approx $500\,MHz.
Because the emissions of a relativistic plasma occur mostly above the plasma frequency in the plasma frame, the maxima of spectral power can be expected to lie above 500\,MHz.

We estimated the height in the magnetosphere for these plasma densities.
After the relativistic transformation of the corresponding plasma densities into the pulsar frame and taking into account a density scaling with the height in the magnetosphere \citep{Goldreich1969,Rahaman2020},
\begin{equation}
        n(r) \approx 5.5\times10^{5} \kappa \left( \frac{1\,\mathrm{s}}{T_\mathrm{pulsar}} \right) \left( \frac{B}{10^{12}\,\mathrm{G}} \right) \left( \frac{R}{r} \right)^3 [\mathrm{cm}^{-3}],
\end{equation}
we can find the emission heights $r \approx 210\,R$ and $r \approx 33\,R$, respectively,
where $\kappa = 10^3$ is the secondary plasma multiplicity factor, $T_\mathrm{pulsar} = 0.25$\,s is the pulsar period, $B = 10^{12}$\,G is the pulsar surface magnetic field, $R = 10$\,km is the neutron star radius, and $r$ is the height in the magnetosphere.


Both simulations were carried out for 140\,000 time steps ($\omega_\mathrm{p} t = 3500$), which allowed a frequency resolution of $\omega/\omega_\mathrm{p} = 2.9 \times 10^{-4}$.
This simulation time is long enough for the streaming instability to cover the instability growth and saturation.
For the bunch interaction, the most intense currents are generated at the beginning of the simulation ($\omega_\mathrm{p}t < 500$).
Then, the currents decreased and saturated.
The bunch interaction was studied for the same number of time steps as the streaming instability to achieve the same resolution in frequency.
The simulation times were $15.6\,\mu\mathrm{s}$ for the relativistic beam simulation and $984\,\mathrm{ns}$ for the bunch interaction simulation, both in the pulsar reference frames.
When we assume that the simulation domain (i.e., the plasma frame) moves at the velocity $c$ along the magnetic field lines in the pulsar reference frame, the domain propagates a distance $\sim$4.7\,km and $\sim$300\,m, respectively, during the simulation time.
In both cases, the distance is smaller than the light cylinder distance, even for a millisecond pulsar ($\sim$50\,km).
On these scales, we neglected the effects of the curved magnetic field lines.

\begin{table}
        \centering
    \caption{Summary of the simulation parameters of the relativistic streaming instability and the plasma bunch interaction in the plasma (simulation) reference frame.}
        \begin{tabular}{lrr}
                \hline \hline
                ~  & Simulations \\
                Parameter & Streaming instability & Bunch interaction \\
                \hline
                Time steps & 140\,000 & 140\,000 \\
                L [$\Delta x$] & 100\,000 & 720\,000 \\
                $\Delta x / d_\mathrm{e}$ & 0.05 & 0.05 \\
                $\omega_\mathrm{p}$ [rad~s$^{-1}$] & $\omega_\mathrm{p} = \sqrt{2}\times 10^{7}$ & $\omega_\mathrm{p} = \sqrt{5} \times 10^{8}$ \\
                $\omega_\mathrm{p} \Delta t$ & 0.025 & 0.025 \\
                $\rho$ & 3.33 & 1 \\
                $n(x)$ & Uniform & Bunch (Eq.~\ref{eq-density}) \\
                $n_0$ [PPC] & $10\,600$ & $2000$ \\
                \hline \hline
        \end{tabular} \\
    \footnotesize{The number of time steps, the simulation lengths in grid cells, the grid cell size, the number of time steps, the plasma frequency, the initial inverse temperature, the initial density profile, and the initial number of particles-per-cell (PPC).}
        \label{tab1}
\end{table}

The plasma particles of species $\mu$ were initialized with a 1D Maxwell-J\"uttner velocity distribution \citep{Juttner1911},
\begin{equation} \label{eq:distr}
        g_\mu(u) = \frac{1}{\gamma_\mathrm{d\mu}}\frac{n_\mu}{2 K_1(\rho_\mu)} \mathrm{e}^{- \rho_\mu \gamma_\mathrm{d\mu} \gamma (1 - \beta \beta_\mathrm{d\mu})},
\end{equation}
\begin{equation}
        \beta = \frac{v}{c} = \frac{u}{c\gamma}, \quad \beta_\mathrm{d\mu} = \frac{v_\mathrm{d\mu}}{c} = \frac{u_\mathrm{d\mu}}{c\gamma_\mathrm{d\mu}},
\end{equation}
where 
$n_\mu$ is the particle density,
$\rho_\mu = m_\mathrm{e}c^2 / k_\mathrm{B}T_\mu$ is the inverse dimensionless temperature,
$m_\mathrm{e}$ is the electron mass, $c$ is the speed of light,
$k_\mathrm{B}$ is the Boltzmann constant, $T_\mu$ is the thermodynamic temperature,
$K_1$ is the MacDonald function of the first order (modified Bessel function of
the second kind),
$\beta$ and $u$ are the particle velocities, 
$\beta_\mathrm{d\mu}$ and $u_\mathrm{d\mu}$ are the species drift velocities,
$\gamma$ and $\gamma_\mathrm{d\mu}$ are the corresponding Lorentz factors. 
The $y$ and $z$ spatial coordinates and associated velocity components are initially zero and are also zero during the simulation because the evolution of the electromagnetic field vector has only nonzero components along the $x$ -axis.
The $x$ and $y$ components were set to zero as any transverse momentum will be radiated away on timescales $\ll \omega_\mathrm{p}^{-1}$.

\subsection{Relativistic beam instability} 
We carried out the streaming instability simulation using four particle species: a background plasma composed of electrons and positrons with a density $n_0 = 10^4$ particles-per-cell (PPC), and a beam composed of electrons and positrons with a density $n_1 = 600$ PPC.
The typical number of macro-particles per Debye length was $2 \times 10^{5}$.
The beam Lorentz factor was $\gamma_\mathrm{b} = 60$, corresponding to the overlapping velocity of the plasma bunches (not to the velocity of the bunch in the magnetosphere) for which we found the instability growth to be highestl \citep{Manthei2021,Benacek2021a}.
From our consideration, the background plasma moves with mean Lorentz factor $\gamma_\mathrm{s} = 100$ in the magnetosphere, and the beam component with a mean Lorentz factor $\gamma_\mathrm{b} \gamma_\mathrm{s} = 6000$.
The range we used covers the typically assumed range of secondary particle Lorentz factors $\gamma = 10^2 - 10^4$ \citep{Arendt2002}.

We assumed a beam-to-background density ratio $r_\mathrm{n} = n_1 / (\gamma_\mathrm{b} n_0) = 10^{-3}$.
The inverse temperature of all species was $\rho_\mu = 3.33$.
Although the inverse temperature was higher than $\rho \approx 1$ found for the secondary plasma by \citet{Arendt2002}, the inverse temperature of the overlapping bunches was higher because the distribution functions become narrower in velocity space when the bunches expand and overlap \citep{Usov2002,Melrose2020a,Benacek2021b}.
The simulation length was $10^5 \Delta$ ($5000\,d_\mathrm{e}$),
which allows a resolution of the wave number $\Delta k c / d_\mathrm{e} = 1.3 \times 10^{-3}$.

Because the plasma has an initially uniform density, the length can in general be shorter or larger.
The simulation length we used represents a good equilibrium of wave-number resolution while keeping computational expenses relatively low.
The simulation length was $106$~km in the plasma reference frame, and
assuming the Lorentz factor $\gamma_\mathrm{s}=100$, the length was $1060$\,m in the pulsar reference frame when the observer is located at $\theta = 0$.

The subtracted net currents are close to zero because the initial drift between electrons and positrons is negligible during the simulation.
Although the net current subtraction does not influence the simulation evolution or the LAE, artifacts might appear in the Fourier space.
To avoid the artifacts, the net current subtraction was initially used, similarly as in the interacting plasma bunches described below.
However, we found the net current subtraction unnecessary for the correct calculation of the LAE.

\subsection{Plasma bunch interaction}
The plasma bunches were simulated by considering two particle species with opposite drift velocities:
electrons with a drift velocity $u_\mathrm{d}/c = 10,$ and
positrons with a negative drift velocity $u_\mathrm{d}/c = -10$.
These drifts can be obtained by accelerating particles during the $\gamma$-ray decay in strong electric fields into electron--positron pairs \citep{Rahaman2020}.
The inverse temperature of both species is $\rho_\mu = 1$ \citep{Arendt2002}.
The initial density profile of the bunch describes the interaction of two consequently emitted bunches.
We covered half of the leading and half of the trailing bunch.

Because the exact density profiles of the bunch along the magnetic field line are now known, we selected an arbitrary smooth function for the initial profile of electrons and positrons as
\begin{equation} \label{eq-density}
        n(x) =   
        \left\{
        \begin{array}{ll}
                0.1 \, n_0, & |x| \leq \frac{l}{2}, \\
                n_0 \, \mathrm{exp}\left\{ - \frac{\left(\frac{L}{2} - |x|\right)^6}{x_0} \right\}     & |x| > \frac{l}{2}, \\
        \end{array}
        \right.
\end{equation}
\begin{equation}
        x_0 = \left| \ln(0.1) \right|^{-\frac{1}{6}} \frac{(L - l)}{2}  \approx 0.870\cdot \frac{L - l}{2},
\end{equation}
where $L = $720,000$\,\Delta$ (36,000$\,d_\mathrm{e}$) is the simulation length,
$l = L/30$ is the distance between bunches,
$n_0$ is the density in the bunch center, in this case, represented by 2000~PPC, and
$x_0$ was chosen such that $n(x)$ be a smooth function at $x = \pm l/2$.
We assumed that the initial plasma density is an even function.

In Eq.~\ref{eq-density}, we chose the generalized Gaussian profile with the power exponent $p=6$ in the expression $\mathrm{e}^{-x^p}$ to make the transition of the bunch and the surrounding plasma steeper and to separate the bunches more from the background plasma than for a Gaussian profile with $p=2$.
For the generalized Gaussian profile, we were inspired by laboratory measurements in electron beam colliders and laser wakefields, which also detected profiles similar to generalized Gaussian with $p>2$ \citep{Decker1995,Papadopoulos2020,Geng2021,Liang2022}.
Nonetheless, the study of how the density profile, for instance, the parameter $p$, specifically influences the properties of the emitted radiation is beyond the scope of this paper.

The typical number of macroparticles per the Debye length along the simulation box is $200-2000$.
The wave number resolution is  $\Delta k d_\mathrm{e} = 1.7 \times 10^{-4}$.
The simulation length is $48.3$~km in the plasma reference frame and $483$\,m in the pulsar reference frame ($\gamma_\mathrm{s} = 100$).
The length in the pulsar frame approximately corresponds to a plasma bunch produced in the polar gap region \citep{Ruderman1975,Ursov1988,Usov2002}.

The net current was subtracted in the simulation.
The current is produced by the relative drift velocity between the electrons and positrons, corresponding to a screening of the Goldreich--Julian magnetospheric currents \citep{Timokhin2010,Timokhin2013}.
If no initial drift is introduced, the bunches may expand, overlap in the phase space, and form the streaming instability above.
In addition, the net current subtraction assumes that the magnetospheric currents do not change significantly in the simulation length scale.

We subtracted the net current in each time step because it leads to the suppression of strong nonphysical aliasing appearing in the $\omega-k$ domain that artificially contributes to the LAE.
Even so, there is no difference in the plasma evolution with and without the net current. 
If the net current is not subtracted, the LAE has a negligible difference for waves at $k=0$.
The difference can be neglected because the net current in the Fourier space has two components with zero or negligible contributions:
(1) The component that is constant in time given by the initial drift between electrons and positrons does not contribute to the emission power because the power in Eq.~\ref{eq:emission_power} is zero for a zero frequency.
(2) The oscillating component in time is produced by averaging local oscillations in the simulation box and does not significantly contribute to the radiation because the average approaches zero at the bunch scale. 
Hence, these waves with $k=0$ contribute negligibly to the emission power.
The obtained LAE is also similar to the net current subtraction.
Moreover, by the current subtraction, we avoided the aliasing in the $\omega-k$ domain.

\section{Numerical model of the linear acceleration emission} \label{methods2}

The LAE was calculated assuming that plasma particles oscillate, generate electric currents, and emit electromagnetic waves.
The theoretical background of the LAE originates from electromagnetic waves emitted by a relativistic oscillating charged particle or a dipole.
Consequently, the approach was generalized to a system of particles characterized by plasma currents \citep{Melrose1991}.
A similar approach is sometimes used by directly exploiting oscillating electric fields \citep[~]{Jackson1998}.
Nonetheless, the electric current and electric field may not be related by a simple relation, as follows from the Maxwell equation, $\nabla \times \mathbf{B} = \mu \left(\mathbf{j} + \epsilon \partial \mathbf{E}/\partial t \right)$, where $\mu$ and $\epsilon$ are the permeability and permittivity of the plasma (not vacuum).

The energy radiated by a plasma region per unit spatial angle $\mathrm{d}\Omega$ and frequency $\mathrm{d}\omega$ that has a current density $\boldsymbol{j}(\boldsymbol{k},\omega)$ is given by \citet[~]{Melrose1991} in SI units
\begin{equation} \label{Eq:A1}
        \frac{\mathrm{d}U(\boldsymbol{k},\omega)}{\mathrm{d}\Omega \mathrm{d}\omega} = \frac{\omega^2}{16\pi^3 \epsilon_0 c^3} | \boldsymbol{n} \times \boldsymbol{j}(\boldsymbol{k}, \omega) |^2,
\end{equation}
where $\omega$ is the frequency, $\boldsymbol{k}$ is the wave vector, and
$\boldsymbol{n} = \boldsymbol{k} / k$ is the unit vector in the direction of the electromagnetic wave vector.
The electric current density $\boldsymbol{j}(\boldsymbol{k}, \omega)$ is represented in Eq.~\ref{Eq:A1} by a Fourier transform over the whole time--space domain,
\begin{equation} \label{eq:current}
        \boldsymbol{j}(\boldsymbol{k}, \omega) = \int \boldsymbol{j}(\boldsymbol{x},t) e^{i\omega t} e^{-i \boldsymbol{k}\cdot\boldsymbol{x}} \: \mathrm{d}t \: \mathrm{d}\boldsymbol{x}. 
\end{equation}

In 1D geometry, in which particles move only along the $x$ direction, and assuming the magnetic field direction along the $x$-axis, the vectors simplify to $\boldsymbol{x} = (x,y,z) = (x,0,0)$, $\boldsymbol{j}(\boldsymbol{x},t) = (j(x,t)\delta(y)\delta(z),0,0)$, where $\delta$ is the Dirac delta distribution.
Thus, the emission is similar to a current-fed antenna.
The currents are located along the $x$-axis (similar to \citep[~]{Jackson1998}).
In this step, we relate the 1D electric currents that infinitely extend in $y-z$ plane in simulations to a region with an extent of $\delta(y)\delta(z)$.

In a real plasma, the emitting plasma can be confined in a region of lateral extent $D$.
The use of $\delta$ functions instead of an arbitrary lateral extent can be considered valid until $\lambda \gg D$, where $\lambda$ is the typical wavelength of the emitted wave.
The emitted waves interfere during their propagation, as follows from the Fourier transform in Eq.~\ref{eq:current}.
If the emission in the perpendicular direction to the magnetic field is at a frequency $\sqrt{2}\times10^7$\,s$^{-1}$ in the plasma frame (as discussed in Appendix~\ref{methods1}), the typical estimated wavelength is $\lambda \approx 133$\,m when we assume a light wave ($\omega / k = c$).
Because the plasma frequency in the emission regions for the bunch interaction is similar to the plasma frequency of the streaming instability, the generated wavelengths are also similar.

To estimate the radiation power in reality, a lateral extent of the currents and volume of the emission region is necessary, even though the lateral dimensions of the electric currents in Eq.~\ref{eq:current} are assumed as $\delta$ functions.
To calculate the power, the electric currents can be considered to form a cylinder along the magnetic field, as presented by \citet[~]{Rylov1978}.
Our results using the $\delta$ functions are self-similar to a case with a cylinder with a cross section of 1\,m$^2$.
The cross section of 1\,m$^2$ is obtained for a cylinder of diameter $D \approx 1.13$\,m, which is still smaller than the estimated wavelengths $\lambda \approx 133$\,m.

Under these assumptions, the Fourier transform of the current density component can be expressed as
\begin{equation} \label{eq:A3}
        j(k, \omega, \theta) = \int j(x,t) e^{i\omega t} e^{-i \hat{k} x} \: \mathrm{d}t \: \mathrm{d}x,
\end{equation}
where $\hat{k} x = \cos(\theta) k x = \boldsymbol{k}\cdot \boldsymbol{x} $ is the projected wave number.
The relation between $k$ and $\hat{k}$ can be understood by that the electric current density with a wave number $\hat{k} = k\cos(\theta)$ produces emission at the wave number $k$ in the direction of the angle $\theta$.

The relation between wave number and frequency is given by the plasma dispersion relation.
Nonetheless, we state as a constraint the refraction index of the plasma, 
\begin{equation} \label{eq:dispersion}
        n = \frac{k^2 c^2}{\omega^2} \equiv 1,
\end{equation}
and obtain $k = \omega/c$.
This is a valid approximation for the vacuum or when the plasma does not significantly influence the propagation of electromagnetic waves.
This situation occurs when the wave frequency is significantly higher than the local plasma frequency.
If the wave frequency approaches the plasma frequency, an additional term appears in the numerator of Eq.~\ref{Eq:A1}, relating the ratio of electric and magnetic energy in the wave \citep{Melrose1986}.
As a consequence, Eq.~\ref{eq:dispersion} considers and restricts that only superluminal waves can contribute to the radiation power because $k' = \frac{\omega}{c}\cos\theta \leq \frac{\omega}{c}$ (where $k' = \frac{\omega}{c}$ is the light line).
This relation was also used by \citet{Melrose2009b}, and it is implicitly assumed by \citet{Reville2010}, for example.

Using the constraint Eq.~\ref{eq:dispersion} in the Fourier transform Eq.~\ref{eq:A3} neglects the radiation by subluminal waves.
For example, the Cherenkov relation, in which the phase velocity is lower than the speed of light, cannot be satisfied by the approach \citep{Melrose2017b}; $k'$ cannot represent the subluminal waves, that is, it cannot be larger than $\frac{\omega}{c}$.
In the pulsar plasma, this relation may be applied for frequencies higher than the local relativistic plasma frequency of the plasma emission region.
It applies when the emission region is surrounded by a diluted plasma that is less dense than in the emission region but not necessarily requires the vacuum.
More details about the limits of this approximation are given in Sect.~\ref{methods2:details}.

The resulting emitted energy is (in the 1D limit)
\begin{eqnarray} \label{eq:emission_power}
        \frac{\mathrm{d}U(\omega,\theta)}{\mathrm{d}\Omega \mathrm{d}\omega} &=& \frac{\omega^2}{16\pi^3\epsilon_0 c^3 } \left| \sin(\theta) j\left( \frac{\omega}{c},\omega,\theta \right) \right|^2.
\end{eqnarray}
The average wave power is defined as
\begin{equation} \label{eq:powerlimit}
                \frac{\mathrm{d}P(\omega,\theta)}{\mathrm{d}\Omega \mathrm{d}\omega} = \lim_{\Delta t \to \infty} \frac{1}{\Delta t} \frac{\mathrm{d}U(\omega,\theta)}{\mathrm{d}\Omega \mathrm{d}\omega}
        ,\end{equation}
where $\Delta t$ is the average time interval.
The time interval corresponds to time over which the electric currents are integrated in the Fourier transform of Eq.~\ref{eq:current}.

Specifically, to obtain the current $j(\vec{k},\omega)$ in Eq.~\ref{eq:current}, the integration limit $\Delta t \rightarrow \infty$ in the Fourier transform was implicitly assumed.
Moreover, because Eq.~\ref{eq:emission_power} contains the currents to the second power, the time interval $\Delta t$ appears in the numerator of the equation.
To show it, we assume that the current
\begin{equation} \label{eq:current_expand}
    j(\vec{k},\omega) = \sum_s j(\vec{k}) 2 \pi \delta(\omega - \omega_s(\vec{k}))
\end{equation}
is a set of waves $s$ with frequencies $\omega_s$ for a given $\vec{k}$.
To obtain the emitted energy in Eq.~\ref{eq:emission_power}, the square of $\delta$ functions can be expanded as \citep[~]{Melrose1991}
\begin{eqnarray}
    [2 \pi \delta (\omega)]^2 &=& 2 \pi \delta(\omega)\delta(\omega) \\
    &=& \int_{-\Delta t/2}^{\Delta t/2} \mathrm{d}t\,\mathrm{e}^{i\omega t} \int_{-\Delta t/2}^{\Delta t/2} \mathrm{d}t' \mathrm{e}^{i\omega t'} \label{eq:innersquare} \\
                              &=& \frac{1}{2} \int_{-\Delta t}^{\Delta t} \mathrm{d}(t-t') \int_{-\Delta t}^{\Delta t} \mathrm{d}(t+t') \mathrm{e}^{i\omega (t + t')} \label{eq:outersquare} \\
                              &=& \Delta t \,2 \pi \delta (\omega).
\end{eqnarray}
To evaluate the $\delta$ functions, the integrals in Eq.~\ref{eq:innersquare} over an inscribed square $-\Delta t /2 < t < \Delta t /2$ and $-\Delta t /2 < t' < \Delta t /2$ in the $t - t'$ plane are calculated in Eq.~\ref{eq:outersquare} using the evaluation over half an outer square.
The square is parameterized by $-\Delta t < t-t' < \Delta t$ and $-\Delta t < t + t' < \Delta t$ and is rotated by $\pi/4$ with respect to the inscribed square.
Moreover, the transformation from the inner to outer square allows us to evaluate the term $\mathrm{e}^{i \omega (t+t')}$ in the second integral over $(t+t')$, while keeping the first integral over $(t-t')$ trivial.
As a result, the emitted energy $U \propto \Delta t$ for $\Delta t \rightarrow \infty$.

Putting Eq.~\ref{eq:current_expand} into Eq.~\ref{eq:emission_power} leads to mutual canceling of $\Delta t$ in \ref{eq:powerlimit}, in agreement with the assumed periodicity over the time interval $\Delta t$.

Therefore, the power $P$ in Eq.~\ref{eq:powerlimit} represents a period-time-averaged electromagnetic power in which individual contributions are coherently added.

For the grid-based current density in Fourier space $j_{i}^l(\omega,k)$ from the simulations (where $i$ and $l$ are grid indices corresponding to spatial and temporal coordinates, respectively), we can numerically calculate the average emission power in the plasma reference frame.
More details about the numerical implementation and the limitations of this approach are stated in the following sections.

\subsection{Implementation details}
\label{methods2:details}
We present several notes on the numerical implementation of the approach described above.
The highest resolved frequency is given by the storage interval  $\Delta t$ of electromagnetic field data in the PIC code. This frequency (together with the wav enumber) then scales to the highest frequency, $\omega_\mathrm{max} = 2 \pi / \Delta t$, in the plasma frame.
We chose a storage time step that was short enough to cover all frequencies presented in this paper well.
        
For the relativistic beam instability simulation, a weak alias appears in currents at a multiple of the plasma frequency as a horizontal superluminal mode in the plasma reference frame.
This cannot be avoided by increasing the size of the simulation domain, the number of particles, or the order of the particle shape function.
Thus, it is also present in some emission properties.
Nevertheless, the region does not significantly influence the finally estimated total flux because its intensity is approximately four orders lower than that of the main superluminal mode.
        
We did not apply any window filter to the space--time array $j_{il}(x,t)$ to avoid aliasing.
In the streaming instability, another weak alias appears at k = 0.5\,m$^{-1}$ and $\omega = (1.5 - 4) \times 10^{7}$\,rad\,s$^{-1}$. 
However, as its currents are at least one order of magnitude lower than the superluminal waves, the relative contribution to the emission power is $\sim$$10^{-2}$.
In the plasma bunch interaction, no significant aliases appear because the most intensive currents are located in the simulation center where the window filter would apply values close to one, that is, not changing the currents.
Moreover, we estimated the effect of possible cosine window filter application of both cases.
The resulting total power varied by a factor of $\lesssim 2$ at most.
                
If electric currents are obtained from PIC simulation instead of individual particles, the calculation profits from the full information obtained by the high-order shape function of the macroparticle.

\subsection{Limits of applicability}
We neglected spatial profiles perpendicular to the magnetic field.
The emission was also provided simultaneously by several sources;
the emitted waves might mutually cancel out by interference.
However, we can assume that the typical emission angle width of one region is significantly smaller than the typical observed half-width  pulse profile of the main pulse.
Hence, each emission source might emit without interference with others into a different angle given by the position of the emission region in the magnetosphere.
Thus, while the observed flux might come from only one coherent source at a given time, the total observed pulse profile is formed by many emission sources.

Although several effects can hinder wave generation and propagation, we assumed that the electromagnetic waves in the studied frequency interval are not modified during their propagation from the emission source to the observer.
This assumption is valid when the radiation propagates through an area that is dominated by a low-density plasma or vacuum.
We also assumed that the source of the emission is denser than the surrounding plasma in the form of a strong inhomogeneity, either filamentary \citep{Urpin2014} or as a longitudinal sheet, resulting from intermittent pair production processes, for instance \citep{Philippov2022}.
In addition, a gap boundary might be suggested close to the slot gap, for example, where the radiation could escape from the slot gap walls into the polar gap region, as was presented by \citet{Philippov2015}, or recently in a PIC simulation of an aligned rotator with higher spatial resolution (\citet{Bransgrove2022}.) 
Fig.~2 in \citet{Bransgrove2022} shows several areas in which the instabilities might lead to radio emission.
Their region (ii), located at the boundary between the open and closed magnetic field lines, might be a candidate for a location in which LAE might occur and from which the radiation could escape.
However, because our simulations resolve kinetic microscales, they cannot simulate a complete magnetosphere in its complexity.

In the following, we consider general situations in which the propagation effects can be neglected.
The situations can be separated into two categories containing (1) absorption effects and (2) a phase delay of the electromagnetic waves because the refractive index is far from one.
For the absorption effects, we assumed that the emission frequency was higher than the absorption edge frequency of the surrounding plasma, which corresponds to its plasma frequency.
For lower frequencies than the surrounding plasma frequency of the bunch, we expect a strong wave absorption.
The small phase delay of the waves requires the plasma refractive index to be close to one in the surroundings of the bunch.
The requirement is fulfilled when the frequencies are significantly higher than the local plasma frequency.
In the dispersion $\omega-k$ space, this effect can be seen as the superluminal electromagnetic waves approaching the light line, which we estimate for frequencies $\omega \gtrsim (1.5-2)\omega_\mathrm{p,loc}$, where $\omega_\mathrm{p,loc}$ is the plasma frequency of the surrounding plasma.

Both assumptions are fulfilled in our considered instabilities when
    (1) the streaming instability or the interacting plasma bunch are formed as secondary particles during a spark event. 
       Their density can exceed the local Goldreich--Julian density, while the surrounding plasma can have a lower density.
       (2) If a plasma bunch is initially created in the lower magnetosphere where the density is higher, it can then propagate to the higher magnetosphere where the surrounding plasma has a lower density and therefore a lower absorption edge frequency.
Moreover, in both cases, the surrounding plasma density might still be high enough to compensate for the Goldreich--Julian currents in the surrounding plasma, and the emitted electromagnetic waves might still propagate similarly to vacuum.

Another option that can decrease the absorption of electromagnetic waves is the relativistic beaming effect.
As the plasma bunch moves at relativistic speeds, the emitted waves are relativistically beamed and propagate in the surrounding plasma in the direction mostly along magnetic field lines.
Because the beamed radiation has a higher frequency than in the plasma frame, the frequency might well be higher than the plasma frequency.
Moreover, an absorption coefficient for these waves is significantly lower because the projected electric component of an electromagnetic wave into the direction of particle motion (along the magnetic field) is small.

An additional possibility is that the emitted electromagnetic waves are initially absorbed by the surrounding plasma. 
However, when the absorbing plasma reaches an arbitrary saturation level of electromagnetic wave energy density, its dispersion properties change, and the wave absorption approaches zero.

In addition to the effects mentioned above, influencing the wave coherence at distances comparable with the size of the emission region, the propagation effects through the magnetosphere were also neglected in our model.
Although there are effects that influence the radio wave propagation, absorption, and polarization \citep{Arons1986,Liubarskii1998,Petrova2002,Petrova2013}, we did not aim to study the effects because their complexity goes beyond the limits of this paper.

\subsection{Tests of implementation}

\begin{figure*}[!ht]
        \includegraphics[width=\textwidth]{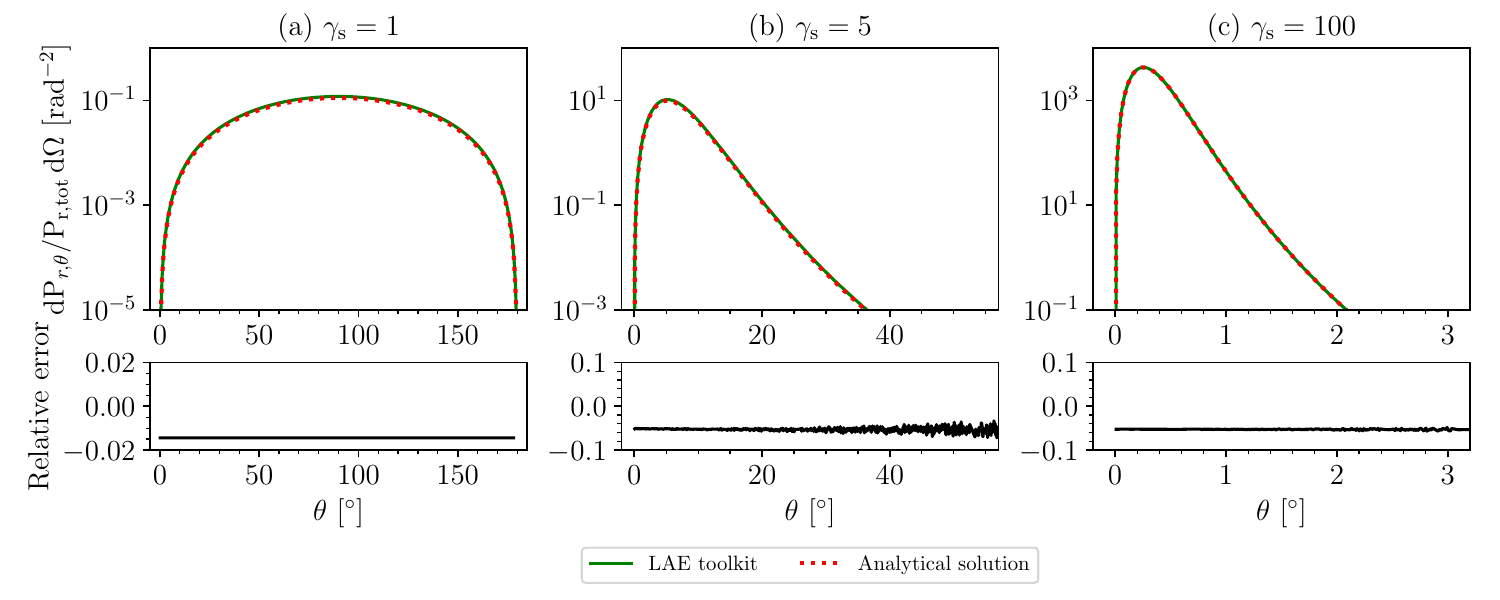}
        \caption{Test of the LAE numerical calculation. Received electromagnetic power per unit spatial angle, normalized to the total received power, and computed for three selected relativistic        transformations $\gamma_\mathrm{s}$ = 1, 5, and 100.
                \textit{Top row:} Received electromagnetic power per unit spatial angle and normalized to the total received power $P_\mathrm{tot}$.
                The analytical solution is represented by Eq.~\ref{eq:emission_anal}.
                \textit{Bottom row:} Corresponding relative errors.
                We assume a point electron oscillating with frequency $\omega_0 = 1$~rad~s$^{-1}$ and
                zero drift velocity that is superposed on a space--time grid of the size $6400\times 6400$ grid cells, and it is inserted as an input into the LAE toolkit.
                The analytical solution is given by Eq.~\ref{eq:emission_anal}.
                The angular resolutions are $1.8\,^{\circ}$ in (a-b) and $0.01\,^{\circ}$ in (c).
                The frequency resolutions are $5\times 10^{-2}\,\mathrm{rad}\,\mathrm{s}^{-1}$ in (a), $5\times 10^{-3}\,\mathrm{rad}\,\mathrm{s}^{-1}$ in (b), and $10^{-2}\,\mathrm{rad}\,\mathrm{s}^{-1}$ in (c).
        }
        \label{fig7}
\end{figure*}

\begin{figure*}[!ht]
        \includegraphics[width=\textwidth]{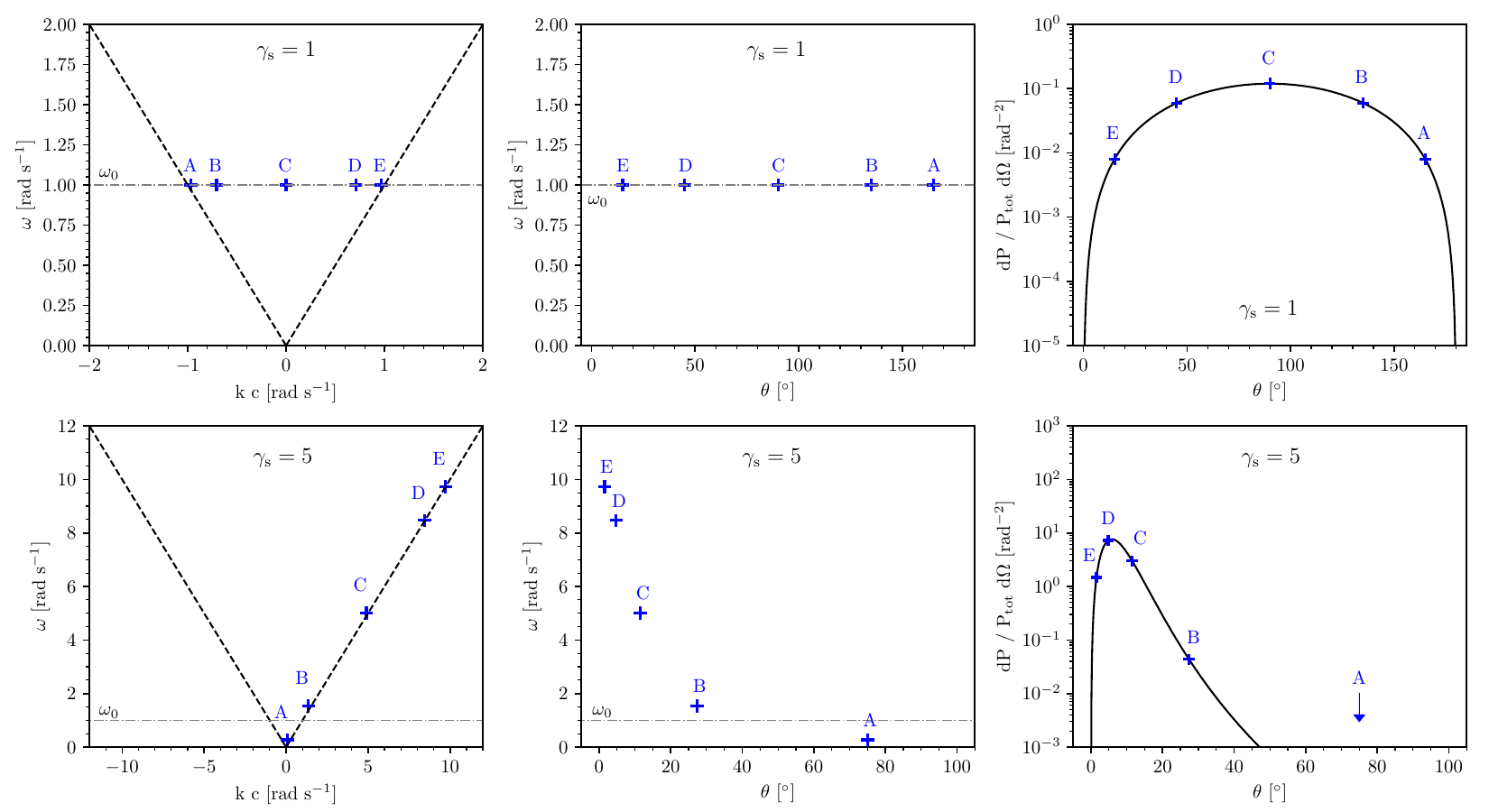}
        \caption{
                Example of how an arbitrarily charged particle can radiate in the plasma frame and in the relativistically shifted frame when approaching an observer.
                We study five arbitrary emitted waves as a function of the frequency, wave number, emission angle, and emission power (per unit spatial angle).
                The particle oscillates at frequency $\omega_0 = 1$~rad~s$^{-1}$ and has an elementary charge.
                $A,B,\ldots,E$ denote five types of emitted waves (blue crosses) with the same frequency $\omega_\mathrm{0}$ and five wave numbers $kc \approx -0.966,-0.707,0,0.707,\text{and }0.966$~rad~s$^{-1}$, both in the plasma reference frame.
                \textit{Top row:} Plasma reference frame.
                \textit{Bottom row:} Relativistically shifted reference frame with $\gamma_\mathrm{s} = 5$.
                \textit{Left column:} Positions of waves in $\omega - k$ space.
                \textit{Middle column:} Positions as a function of frequency and their emission angle.
                \textit{Right column:} Position denoted by arrows at the analytical emission function.
                \textit{Dashed black line:} Light line $\omega = k c$.
                \textit{Dash-dotted gray line:} Frequencies $\omega_\mathrm{0}$.
                \textit{Black line:} Analytically estimated electromagnetic emission by oscillating particle from Eq.~\ref{eq:emission_power}.
        }
        \label{fig8}
\end{figure*}

We tested the software toolkit in a one-particle approximation and for an oscillating coherent wave with an arbitrary spatial size.
We assumed that an electron oscillates with frequency $\omega_0 = 1\,\mathrm{rad}\,\mathrm{s}^{-1}$ for nonrelativistic to relativistic drift velocities $\beta = 10^{-3}$ throughout the magnetic field.
Specifically, the size of the particle orbit is negligible in comparison with the size of one grid cell.
The drift velocity we used corresponds to the average particle velocity over many orbits in the observer reference frame.
Thus, it recalls an oscillating dipole.
As follows from Eq.~\ref{Eq:A1}, there is no feedback of particle emission on its motion.

We superposed the currents onto a grid and input them into the LAE toolkit instead of the simulation currents.
Independently, we calculated received power as a function of the emission angle using
\begin{equation} \label{eq:emission_anal}
        \frac{dP_\mathrm{r,\theta}}{d\Omega} = \frac{q^2}{16 \pi^2 \epsilon_0 c} \left\langle \frac{\left[ \boldsymbol{n} \times \left((\boldsymbol{n}-\boldsymbol{\beta})\times \dot{\boldsymbol{\beta}} \right) \right]^2}{(1 - \boldsymbol{\beta}\cdot \boldsymbol{n})^6} \right\rangle_t,
\end{equation}
where $q$ is the charge of the particle, and $\epsilon_0$ is the permittivity of vacuum,
$\boldsymbol{n}$ is the direction vector of the emission, $\boldsymbol{\beta}$ is the particle velocity, $\boldsymbol{\dot{\beta}}$ is the particle acceleration, both normalized to the light speed, and $\langle \rangle_t$ denotes time average over the particle oscillation.
We tracked the particle for 20 oscillation periods in 2000 time steps.

The resulting electromagnetic received power obtained by the LAE toolkit agrees well with the analytical solutions of emission power as a function of emission angle Eq.~\ref{eq:emission_anal} (see Fig.~\ref{fig7}).
For large $\gamma_\mathrm{s}$, they manifest a systematic error of $\approx 5$~\%.
We also tested a particle with a spatial length (current wave) represented by a Gaussian shape function.
The emission angle narrows with increasing the particle length, and it approaches a delta distribution for an emission length going to infinity.

Figure~\ref{fig8} shows an example of how the electron emission can be represented in dependence on frequency, wave number, emission angle, and emission power into a unit spatial angle.
Two reference frames are considered: the plasma frame and a relativistically shifted frame with $\gamma_\mathrm{s} = 5$.
We assumed that the electron source approaches the observer.
The emission angle $\theta$ is measured from the magnetic field direction, which is also the oscillation axis.
Five arbitrary waves $A$, $B$, \ldots, $E$ were selected.
Their frequency is $\omega_0 = 1\,\mathrm{rad}\,\mathrm{s}^{-1}$ , and the five wave numbers are $kc \approx -0.966,-0.7,07,0,0.707,$ and $0.966$~rad~s$^{-1}$, corresponding to the emission angles $165^{\circ},135^{\circ},90^{\circ},45^{\circ},$ and $15^{\circ}$ in the plasma reference frame.
For example, point $A$ shifts to a lower frequency after the relativistic transformation compared to its position in the plasma frame.
Nevertheless, its emission power is negligible compared to the emission power of other waves.
Waves $B$--$E$ significantly increase their frequency after relativistic transformation.
While most emission power comes from the wave $C$ in the plasma frame,
most of the power is represented by the wave $D'$ in the relativistic reference frame.
These wave representations and transformations are helpful for intuitively understanding our results.

\subsection{Localization of electric currents} 

\begin{figure*}
        \centering
        \includegraphics[width=0.45\textwidth]{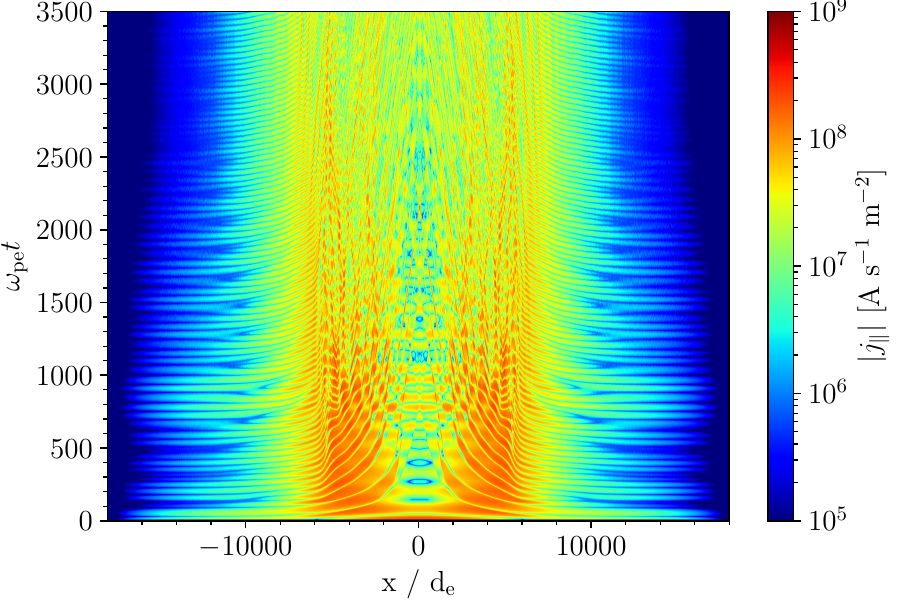}
        \caption{
                Time evolution of the electric current density in the plasma bunch interaction along the simulation domain.
                The currents are primarily localized in the simulation center and decrease at the boundaries.
    Therefore, the currents close to the simulation boundaries do not significantly contribute to the LAE.
        }
        \label{fig10}
\end{figure*}

\begin{figure*}
        \centering
        \includegraphics[width=0.9\textwidth]{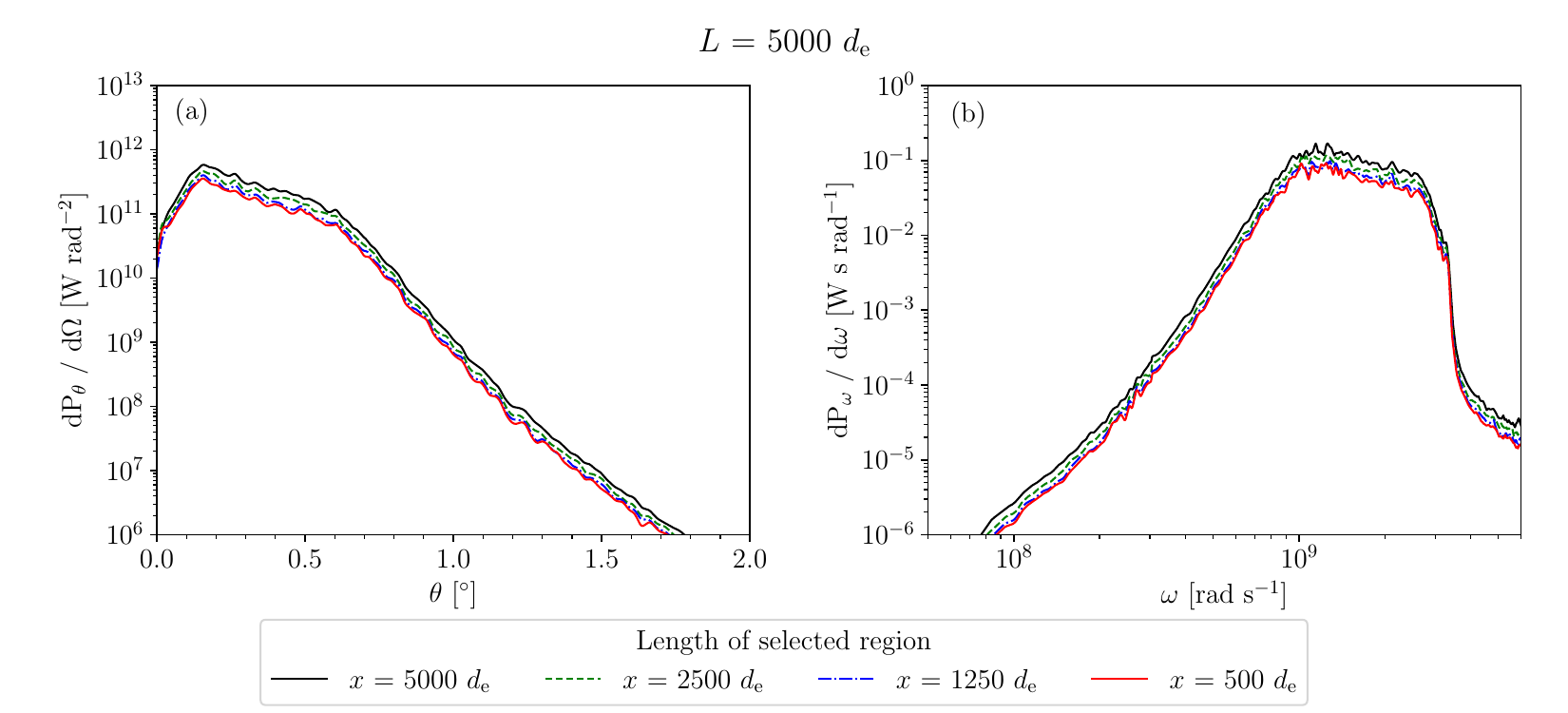}
        \caption{
                        Received radiation power as a function of (a) the angle and (b) the frequency for varying sizes $x$ of selected subdomains from the simulation domain of a length $L = 5000\,d_\mathrm{e}$.
        The selected subdomain is always positioned in the center of the simulation domain.
        }
        \label{fig11}
\end{figure*}

One of the conditions for the emitted energy in Eq.~\ref{eq:emission_power}  is that the electric currents must be localized in the analyzed region \citep{Jackson1998}. In this section, we show how the localization of currents can be addressed in the two simulations from Appendix~\ref{methods1}.
In general, the simulations have no net current because the immediate net current is calculated and subtracted at each time step.
Below, we analyze how the localization of the emission regions is reached in both simulations.

\paragraph{Plasma bunch interaction} 
 The evolution of the electric current density throughout the simulation domain is shown in Fig.~\ref{fig10}.
The currents are localized close to the simulation center, and their amplitudes approach zero close to the simulation boundaries, where they are at least four orders of magnitude smaller than in the simulation center.
Therefore, these regions produce an emission power that is approximately eight orders of magnitude lower, $P \sim j^2$, and we denote most of the currents as localized.

\paragraph{Streaming instability} 
The situation is different from the case above.
The electric currents do not decrease close to the simulation boundary.
They are approximately uniformly distributed throughout the simulation domain.
However, we can select a subdomain of the whole simulation domain, whose boundaries can be considered as not periodic, and estimate the LAE properties in the subdomain.
The subdomain can be regarded as an independent simulation with real boundary conditions.
If the emission of the whole domain is similar to the emission of any subdomain, we can assume that the implemented method produces the same emission power for the periodic as well as the nonperiodic boundary conditions.

To show how the emission varies between the whole domain and its subdomains, we show the received power as a function of the angle and frequency in Fig.~\ref{fig11}.
The total size of the simulation domain is $L=5000\,d_\mathrm{e}$, and we selected subdomains of sizes $l=2500, 1250,$ and $500\,d_\mathrm{e}$.
The subdomains were selected so that each subdomain was centered in the original domain.

The emission power produced by the whole simulation domain has similar profiles as those produced by the subdomains.
The emission power of the smallest subdomain is systematically $\sim 2.5$ times lower than the power of the whole domain.
Hence, the results regarding the emission power from the whole simulation domain can be taken into account, and the power is similar to the nonperiodic domains with localized currents.
Based on this analysis, the calculated power can represent the emission power by the studied instabilities well.

\section{Relativistic beaming} \label{methods3}
There are more ways to calculate the wave power in the pulsar frame, but we calculate the emission power in the plasma frame in Appendix~\ref{methods2} and transform it relativistically into the pulsar (observer) frame.

Denoting the quantities in the plasma (simulation) reference frame by primes and in the pulsar (observer) reference frame without primes, the wave power is converted as \citep[ ]{Rybicki1986}
\begin{equation} \label{eq:relativistic-transform}
    \frac{\mathrm{d} P}{\mathrm{d} \Omega \mathrm{d} \omega} = \mathbf{D}_\mathrm{D}^3 \frac{\mathrm{d} P'}{\mathrm{d} \Omega' \mathrm{d} \omega'},
\end{equation}
assuming that the frequency scales as $\mathrm{d}\omega = \mathbf{D}_\mathrm{D} \mathrm{d}\omega'$, where $\mathbf{D}_\mathrm{D} = 1 / \gamma_\mathrm{s}(1 - \beta_\mathrm{s} \cos(\theta))$ is the relativistic Doppler shift, and $\gamma_\mathrm{s} = (1 - \beta_\mathrm{s}^2)^{-\frac{1}{2}}$ is the transformation Lorentz factor.
We assumed a negative value $\beta_\mathrm{s}$ for a plasma approaching the observer.

We note a substantial difference between the total power emitted by a particle (or a coherent plasma region) and the total received power by an observer.
The \textit{total emitted power} is the same in all reference frames.
For example \citep[~]{Griffiths2017}, the total instantaneous emitted power by a particle undergoing an acceleration $a_\parallel$ along a magnetic field line is expressed as
\begin{equation}
        P = \frac{q^2}{6\pi \epsilon_0 c^3} \gamma^6 a_\parallel^2,
\end{equation}
where $q$ is the particle charge, $\gamma = (1 - v_\parallel^2/c^2)^{-\frac{1}{2}}$ is the particle Lorentz factor, and $v_\parallel$ is the particle velocity along the magnetic field.
It may seem from Eq.~\ref{eq:relativistic-transform} that the total power increases with the particle $\gamma^6$ factor, but the total emitted power does not depend on the Lorentz factor because the acceleration is relativistically transformed as $a_\parallel \sim 1/\gamma^3$.
Thus, the particle emits the same power in all relativistic reference frames.
In contrast to the total \textit{emitted} power, the total \textit{received} power scales with the relativistic factor as $\sim \gamma_\mathrm{s}^2$ \cite[Chapter~4.8]{Rybicki1986}.
The difference is in the time interval used to estimate the power.
While the emitted power is associated with the time interval in which an emission of a given energy occurs, the received power is measured in the time interval of measurement by a stationary observer.
Hence, the scaling as $\sim \gamma_\mathrm{s}^2$ can be obtained by integrating Eq.~\ref{eq:relativistic-transform} over the spatial angle and frequency.

From this definition, there is no difference between the total emitted and total received power in the plasma reference frame because $\gamma_\mathrm{s} = 1$.
However, in the pulsar frame, the total received power is higher than the emitted power by a factor of $\sim \gamma_\mathrm{s}^2$ for $\gamma_\mathrm{s} \gg 1$.
The considered emission power in the results is the strictly received power.

\end{appendix}

\end{document}